\documentclass[]{acmart}

\usepackage{amsmath,amsfonts}
\usepackage{algpseudocode}
\usepackage{algorithm}

\usepackage{amsmath}
\usepackage{multirow}
\algnewcommand\algorithmicforeach{\textbf{for each:}}
\algnewcommand\ForEach{\item[ \algorithmicforeach]}

\usepackage{multicol}
\usepackage{graphicx}
\usepackage{xcolor}
\usepackage{textcomp}
\usepackage{enumitem}
\usepackage{soul}
\usepackage{booktabs,tabularx,enumitem,ragged2e}
\usepackage{xcolor,colortbl}

\AtBeginDocument{%
  \providecommand\BibTeX{{%
    \normalfont B\kern-0.5em{\scshape i\kern-0.25em b}\kern-0.8em\TeX}}}

\setcopyright{acmcopyright}
\copyrightyear{2018}
\acmYear{2018}
\acmDOI{XXXXXXX.XXXXXXX}

\acmPrice{15.00}
\acmISBN{978-1-4503-XXXX-X/18/06}

\begin{document}


\title{The $\mathrm{ACAC_{D}}$ Model for Mutable Activity Control and Chain of Dependencies in Smart and Collaborative Systems}


\author{Tanjila Mawla}
\affiliation{%
  \institution{Department of Computer Science, Tennessee Tech University} \city{Cookeville} \state{TN} \country{USA}
}
\email{tmawla@tntech.edu}

\author{Maanak Gupta}
\affiliation{%
  \institution{Department of Computer Science,  Tennessee Tech University}\city{Cookeville} \state{TN} \country{USA}
}
\email{mgupta@tntech.edu}


\author{Safwa Ameer}
\affiliation{%
  \institution{Institute for Cyber Security (ICS) and  NSF C-SPECC Center, University of Texas at San Antonio} \city{San Antonio} \state{TX} \country{USA}
}
\email{safwa.ameer@utsa.edu}

\author{Ravi Sandhu}
\affiliation{%
  \institution{Institute for Cyber Security (ICS) and  NSF C-SPECC Center, University of Texas at San Antonio} \state{TX} \country{USA}
}
\email{ravi.sandhu@utsa.edu}


\renewcommand{\shortauthors}{Mawla et al.}
\begin{abstract}
With the integration of connected devices, artificial intelligence, and heterogeneous networks in IoT-driven cyber-physical systems, our society is evolving as a smart, automated, and connected community. In such dynamic and distributed environments, various operations are carried out considering different contextual factors to support the automation of collaborative devices and systems. These devices often perform long-lived operations or tasks (referred to as activities) to fulfill larger goals in the collaborative environment. These activities are usually mutable (change states) and interdependent. They can influence the execution of other activities in the ecosystem, requiring \textit{active} and real-time monitoring of the entire connected environment. Traditional access control models are designed to take authorization decisions at the time of access request and do not fit well in dynamic and collaborative environments which require continuous active checks on dependent and mutable activities.

Recently, a vision for activity-centric access control (ACAC) was proposed to enable security modeling and enforcement from the perspective and abstraction of interdependent activities. The proposed ACAC incorporates four decision parameters: Authorizations (A), oBligations (B), Conditions (C), and activity Dependencies (D) for an \textit{object agnostic} continuous access control in smart systems. In this paper, we take a step further towards maturing ACAC by focusing on activity dependencies (D) and developing a family of formal mathematically grounded models, referred to as $\mathrm{ACAC_{D}}$. These formal models consider the real-time mutability of activities as the critical factor in resolving \textit{active} dependencies among various activities in the ecosystem.
Activity dependencies can form a chain where it is possible to have dependencies of dependencies. In ACAC, we also consider the chain of dependencies while handling the mutability of an activity. We highlight the challenges (such as multiple dependency paths, race conditions, circular dependencies, and deadlocks) while dealing with a chain of dependencies, and provide solutions to resolve these challenges. We also present a proof of concept implementation of our proposed $\mathrm{ACAC_{D}}$ models with performance analysis for a smart farming use case. This paper addresses the formal models' intended behavior while supporting activities' dependencies. Specifically, it focuses on developing and categorizing mathematically grounded activity dependencies into various ACAC sub-models without formal policy specification and analysis of theoretical complexities, which are intentionally kept out of the scope of this work.

\end{abstract}


\maketitle


\vspace{-3mm}
\section{Introduction}
Internet-of-Things (IoT) is a rapidly growing technology integrating billions of connected devices and artificial intelligence over heterogeneous networks, facilitating smart and collaborative ecosystems such as smart farming, smart manufacturing, smart cars, and e-health monitoring. In such dynamic and distributed environments, data-driven applications are widely used. Thousands of devices collect and utilize data from users, devices, and environments to support automation collaboratively. The ultimate goal of a futuristic community is to establish an autonomous smart ecosystem for  human-driven domains where everything is connected, continuously communicating, sharing information, and triggering actions.

However, ensuring efficiency and accuracy for such systems while addressing growing security and privacy issues raises serious challenges in these smart communities' operational and administrative aspects. With increasing number of connected and interacting devices, 
the attack surface in such systems is continuously expanding. 
While cybersecurity is a top national priority and much progress has been made to ensure protection from cyber-attacks, IoT-driven smart systems security raises a host of new challenges. The convergence of
the physical and cyber world introduces new automated attack dimensions which are hard to analyze, and engender substantial risk in maintaining the integrity of physical and cyber resources.
Significant challenges to secure connected and IoT-driven systems include threat modeling, proposing mathematically grounded fundamental security approaches, continuous vulnerability assessment, and
designing adaptable autonomous defense mechanisms to thwart rapidly evolving cyber-physical threats in this growing, connected, collaborative, and distributed ecosystem. These systems demand real-time \textbf{active} monitoring of operations and \textit{activities} with the contextual information of multiple device states and environmental conditions for continuous authorization and system security.
Access control solutions are extensively used to secure computer systems from unwanted and unauthorized access. Several traditional and extended access control solutions using discretionary, mandatory, role-based, or attribute-based approaches have been proposed to offer security needs for smart and connected systems \cite{EGRBAC,schuster2018situational,capBac,gupta2020access, gupta2020secure, HABAC,thakare2020parbac,bhatt2021attribute,gupta2019dynamic, FedCAC,extendedACON, cathey2021edge, colombo2021regulating}. However, traditional access control systems fall short in terms of dynamicity, scalability, mutability, and real-time monitoring needs of smart ecosystems.  
As we approach towards a fully automated, coordinated, data-driven, and highly connected future community supporting multi-domain/administered distributed collaborative devices, we need \textit{active} access control models which can adapt to the dynamic context of the ecosystem, continuously monitor the changing access permissions and activities, and handle device failures while ensuring safety and security of the system. 

In response, recently, Gupta and Sandhu \cite{gupta2021towards} proposed a novel activity-centric access control (ACAC) paradigm supporting \textit{activity} as the fundamental abstraction for the active run-time management of security in smart and collaborative systems. 
Intuitively an \textit{activity} is a long-lived continuous event performed by a device in an automated system. Further, these activities change states as they progress and are also inter-dependent, i.e. an activity can control the execution of other activities in the ecosystem. In addition, these activities have chain of dependencies, meaning, an activity A is dependent on activity B, which in-turn is dependent on activity C, referred as\textit{ dependencies of dependencies}.  Our previous work \cite{mawla2022bluesky} proposed the integration of four decision parameters \textit{Authorizations} (A), \textit{oBligations} (B), \textit{Conditions} (C) and \textit{Dependencies} (D) in ACAC, as discussed in Section \ref{sec:motivate}. Further, since smart systems have thousands of connected devices and frequent device failures, it is inefficient for a subject to decide (while making an access request) which particular device will perform the requested activity.
In such cases, it is critical to shift to an \textit{object-agnostic} model, where the system decides which object\footnote{Since, an activity is typically performed by an IoT device in smart ecosystems, we treat the terms object and device as equivalent in activity-centric access control.} is \textit{best} to perform the activity, considering dependencies and other decision factors. This \textit{object-agnostic} approach is very relevant in dynamic and scalable smart ecosystems where devices are randomly added or removed as the system scales. The goal is to approach security modeling and enforcement from the perspective (and abstraction) of activities and their dependencies in collaborative systems.

In this work, we propose a formal mathematically grounded family of ACAC models for activity dependencies (D), referred to as $\mathrm{ACAC_{D}}$. We also show how these models can accommodate the chain of dependent activities providing solutions to some open problems.
The main contributions of this paper are as follows.
\begin{itemize}[leftmargin=*]
  \item We motivate the need for \textit{object-agnostic} access control which supports the mutability of dependent activities. We highlight the limitations of the existing access control models and distinguish ACAC in terms of dynamic activity dependencies, scalability, and activity mutability.
  \item We investigate the activity dependencies (D) component of the ACAC model. Toward this, we propose a family of six $\mathrm{ACAC_{D}}$ sub-models that cover pre-, post-, and ongoing dependencies.
  \item We provide formal definitions for $\mathrm{ACAC_{D}}$ sub-models and illustrate their intended behavior under different dependencies.
  \item We investigate and analyze the chain of dependencies for a requested activity in different stages of its life cycle. We highlight the challenges of resolving the chain of dependencies and propose solutions.
  \item We demonstrate $\mathrm{ACAC_{D}}$ sub-models with use case scenarios (including chain of dependencies) and present a proof of concept implementation to illustrate its application using commercially available technologies.
\end{itemize}

The rest of the paper is as follows. Section \ref{sec:motivate} motivates the need for activity-centric model, discusses the relevant background, and highlights the limitations of existing access control models. Section \ref{Formal_ACAC} presents our proposed family of $\mathrm{ACAC_{D}}$ models with example use cases. Section \ref{chain_of_dependencies} illustrates the challenges while resolving chain of dependencies and show how combination of $\mathrm{ACAC_{D}}$ sub-models are used to resolve a chain of dependencies. Section \ref{sec:implementation} provides a prototype implementation of $\mathrm{ACAC_{D}}$ models and evaluates the performance with comprehensive smart farming use case. Section \ref{related_work} discusses relevant literature on access control models and background. Section \ref{conclusion} concludes the paper.



\section{Motivation for Activity-Centric "Active" Access Control}
\label{sec:motivate}
In smart and collaborative ecosystems, an activity is referred to as a long-lived continuous task that is performed by a device.
At any given moment, thousands of activities and operations could be carried out depending on the workflow needs while considering related and different contextual factors. Activities in such systems are inter-dependent and can constrain the execution of each other. By an "Active" access control model for activity control, we refer to
a security approach enforcing access control requirements where
the system administrator or an automated system constantly monitors workflow needs, the state of the activity, and the decision (to initiate, continue, hold or revoke an activity) parameters. These decision parameters consist of authorizations, obligations, conditions, and dependencies on other activities.
A user,  device, or environmental event can request an activity based on the system workflow and efficiency needs. In general, the most suitable device can be assigned based on the decision parameters to satisfy the activity request.


\begin{figure}[!t]
    \centering
    \includegraphics[scale=0.35]{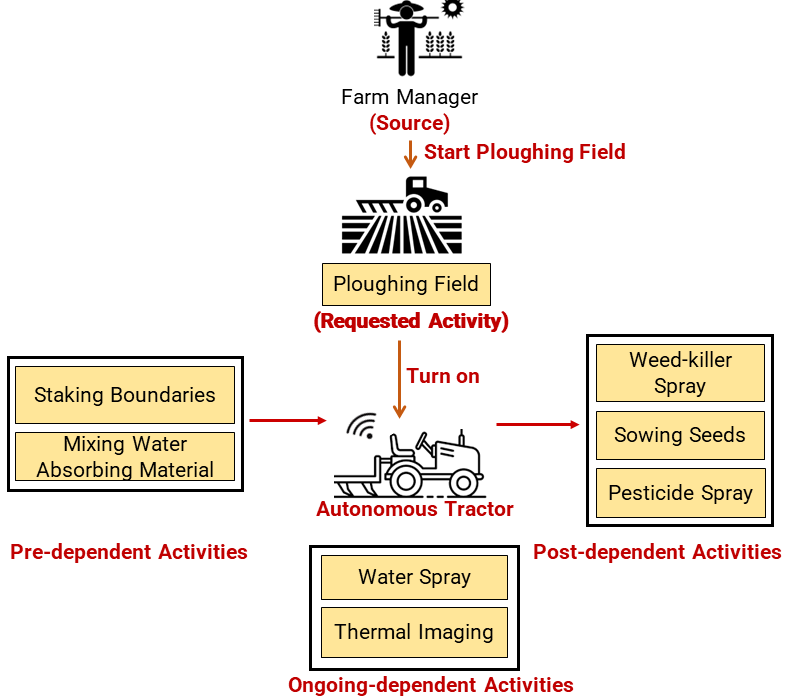}
    \centering
    \caption{Dependencies among Activities.}
    \vspace{-6mm}
    \label{dependent_activities}
\end{figure}
In the example scenario shown in Figure \ref{dependent_activities}, an activity, \textit{ploughing field} is requested by a user, \textit{farm manager}. The system finds the most suitable device, which in our case is the \textit{autonomous tractor}, to perform this requested activity. The corresponding operation, \textit{turn-on} (calculated by the system based on the requested activity and selected device), is performed (if all decision parameters are satisfied) on behalf of the requesting source to initiate the activity, \textit{ploughing field}. However, whether the request is allowed or denied depends on the contextual information, including resolving the dependencies on various other activities in the system. As shown in the figure, there could be three sets of dependent activities; \textit{pre-dependent }, \textit{ongoing-dependent}, and \textit{post-dependent}. Pre-dependent activities are checked before allowing the requested activity, ongoing-dependent activities are checked to ensure whether the execution of the requested activity can be continued or not (if dependencies are violated), and post-dependent activities are checked after the requested activity is revoked, on hold or finished. 
In this example, the requested activity  \textit{ploughing field} can be allowed only if the pre-dependent activities (\textit{staking Boundaries, mixing water absorbing material}) are already running. 
The continuity of the execution of the requested activity depends on the state of ongoing-dependent activities (\textit{water spray} and \textit{thermal imaging}). Finally, different post-dependent activities (\textit{weed-killer spray, sowing seeds, pesticide spray}) are checked after the \textit{ploughing field} activity is finished. The activities are \textbf{mutable} in nature, and can change their states (discussed in Section \ref{Formal_ACAC}) to fulfill the dependency requirements. For example, an activity control policy can be that the \textit{water spray} must be inactive while \textit{ploughing field} is running. In such case, if \textit{water spray} is running, it needs to transition to the finished or revoked state to ensure that it will be inactive immediately (if there is no post-dependent activity) to continue the activity, \textit{ploughing field}. Clearly, this approach requires continuous monitoring and real-time \textbf{active} dependency checks, making the ACAC novel and relevant for smart and collaborative ecosystems.




\begin{figure}
    \centering
    \includegraphics[scale=0.4]{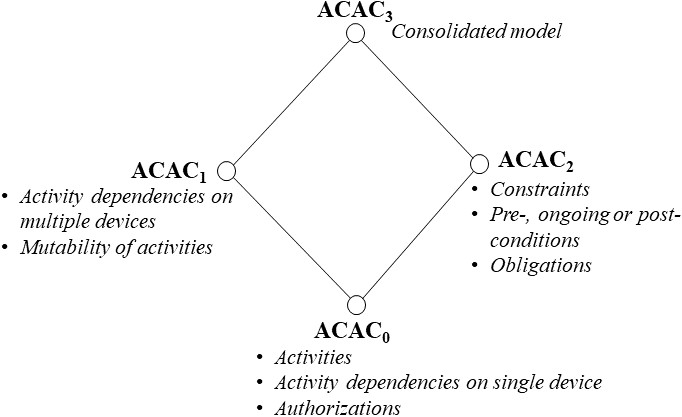}
    \centering
    \caption{A Framework for a Hierarchy of ACAC models \cite{mawla2022bluesky}}
    \vspace{-5mm}
    \label{fig:family}
\end{figure}



Recently, Mawla et al  \cite{mawla2022bluesky} proposed the components of the ACAC model and an incremental approach in a hierarchical framework to fully mature activity-centric access control. Instead of a monolithic model, different features are gradually added to a family of ACAC models, as illustrated in Figure \ref{fig:family}.
The fundamental concept of activity and activity dependencies on a single device is captured in $\mathrm{ACAC_{0}}$. In $\mathrm{ACAC_{1}}$, activity dependencies on multiple devices and the mutability of activities are addressed. Note that, the activity dependencies on single or multiple devices are immaterial as ACAC is an \textit{object-agnostic} model and considers security modeling at the \textit{activity} abstraction. Therefore, both scenarios can be captured in $\mathrm{ACAC_{1}}$ and the most suitable device is automatically decided by the system based on different factors. $\mathrm{ACAC_{2}}$ adds static and dynamic constraints on activities, conditions (including system or environmental, e.g., weather, location), usage count, and obligations (required actions by the source). $\mathrm{ACAC_{3}}$ is built on top of all ACAC models, which is the consolidated and detailed model to implement activity decision control in smart systems. Clearly, $\mathrm{ACAC_{3}}$ will eventually cover the Authorizations (A), oBligations (B), Conditions (C) and Dependencies (D), as decision parameters, and can also be referred as $\mathrm{ACAC_{ABCD}}$.

However, in this paper, we focus on the activity dependencies (D) component of ACAC. We develop formal mathematically grounded models for $\mathrm{ACAC_{D}}$, which support the activity dependencies on multiple devices and the mutability of activities. We investigate the dependencies of dependencies to generate more fine-grained access control model. We also present a prototype implementation of our proposed family of $\mathrm{ACAC_{D}}$ models and evaluate them using a comprehensive smart farming use case scenario with multiple activity requests and activity dependencies along with chain of dependencies.

\subsection{Threat Model}

\begin{figure}[!htb]
    \centering
   \includegraphics[width=\textwidth]{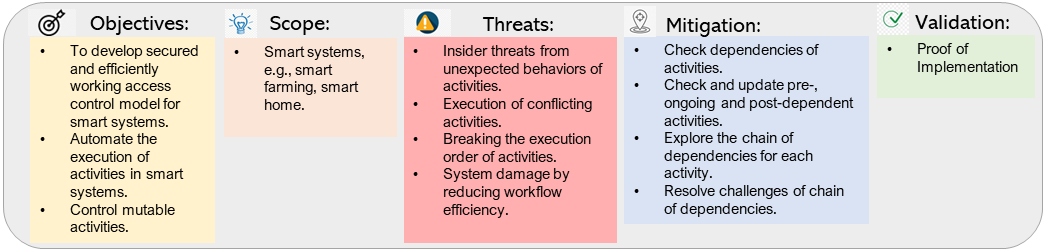}
    \centering
    \caption{Threat Model.}
    \label{fig-threat_model}
\end{figure}
Figure \ref{fig-threat_model} represents the threat model of our proposed $\mathrm{ACAC_{D}}$ model. This model is proposed based on activity dependencies in smart IoT-based systems where safety and security are the major concerns during the automation of different activities. Note that, the model acknowledges the presence of both immutable and mutable activities. Existing threats can exploit the vulnerabilities while the system wants to control the mutable activities according to the workflow preserving the safety of the system. In smart and connected systems, attacks can occur intentionally or accidentally by exploiting known and unknown vulnerabilities. Adversaries can be insiders or outsiders. Our primary emphasis is on insider threats that arise from unexpected behaviors, which can compromise system safety, violate workflows, and hinder efficiency. In complex systems with multiple devices performing various activities, a requester may not have knowledge of all the activities occurring. Consequently, simply checking authorization is insufficient for making activity decisions, as authorized users may still be restricted by activity dependencies. 
By considering these dependencies, we ensure the safety and security of the the system from conflicting activities, disruptions to the execution order, and violations of usage rules. Additionally, this approach enables the execution of emergency and high-priority activities. In our approach, denoted as $\mathrm{ACAC_{D}}$, we assume that all sources are authorized for the requested activity, with a focus on verifying activity dependencies. We also take into account resolving the dependency chains to fulfill activity requests efficiently and in a secured way considering existing threats. However, we acknowledge the challenges involved in resolving these dependency chains and propose mitigation techniques. Our implementation serves as proof of the robustness of the $\mathrm{ACAC_{D}}$ model.

\subsection{Distinction from existing access control}

In access control literature, different models (beyond classical DAC, MAC, and RBAC) have been proposed considering various decision parameters. 
Detailed in work by Mawla et al. \cite{mawla2022bluesky}, in this subsection, we review some of the closely related models with the ACAC model, Task-based Authorization Controls (TBAC) \cite{TBAC},  Usage Control (UCON) \cite{park2004uconabc},  Activity-Centric Access Control for social computing (ACON) \cite{ACON},  Attribute-based Access Control (ABAC) \cite{ABAC, ABACGupta, bhatt2021attribute, bhatt2020abac}, and highlight key distinguishing features. 


Table \ref{comparison} summarizes the distinguishing features which are most relevant in terms of the notion of activity and activity-dependencies between ACAC and other models. The first column in the table contains the name of the models. The rest of the columns mention the key distinguishing features (we selected five, but could be more) among these models and if the models support these keys (Yes) or not (No). 
The key factors are \textit{abstraction of activity}, \textit{dynamic activity dependencies} (meaning activities are inter-dependent  and dynamically calculated based on different factors), \textit{object-agnostic} (refers that corresponding object for an activity will be decided by the system rather than by the requesting source at the time of request), \textit{dependent activity mutability} (the property of changing dependent activity states), and \textit{ongoing monitoring of the system context} (the system context information such as dependencies, usage, environmental conditions, etc., are continuously evaluated to support context-based access decisions).

\begin{table}[t]
\setlength{\tabcolsep}{4pt}
\renewcommand{\arraystretch}{1}
\centering
\caption{Comparison of Features Proposed in ACAC Model}
\vspace{-2mm}
\label{tab5:sum}
\scalebox{0.7}{%
\begin{tabular}{*{30}{|p{1.67cm}|p{1.5cm}|p{1.6cm}|p{1.67cm}|p{1.6cm}}}
\hline

\textbf{Access Control Models} & \textbf{Abstraction of activity}&\textbf{Dynamic activity dependencies}& \textbf{Object-agnostic}&  \textbf{Dependent activity mutability}& \textbf{Ongoing monitoring of system context}\\
\hline
TBAC&Yes&No&No&No&No
\\
\rowcolor{lightgray!40!}
\hline
UCON&No&No&No&No&Yes
 \\
 \hline
ACON&Yes&No&No&No&No
 \\
  \rowcolor{lightgray!40!}
 \hline
ABAC&No&No&No&No&No
 \\
 \hline
\textbf{ACAC}&YES&YES&YES&YES&YES
 \\
\hline

\end{tabular} 

}
\label{comparison}
\vspace{-4mm}
\end{table}

\textbf{Distinction from UCON: } The proposed $\mathrm{ACAC_{ABCD}}$ model is inspired by the UCON \cite{UCONConf, park2004uconabc}. However, there are significant distinctions between $\mathrm{UCON_{ABC}}$ and $\mathrm{ACAC_{ABCD}}$ models. 
UCON supports attributes' mutability which is different from activity mutability supported by ACAC. UCON, primarily designed for digital rights management, does not have a notion of activity (which is a prolonged state of a device). In addition, UCON defines the object on which the operation is requested, which is different than $\mathrm{ACAC_{ABCD}}$, which is an \textit{object-agnostic} model. Further, the chain of dependencies supported in $\mathrm{ACAC_{ABCD}}$ is not considered in UCON. The dependencies in ACAC can be on the same or different objects. Where the activity is actually executing or which source started the activity is irrelevant. The abstraction of activity in ACAC makes it easier to manage connected systems in terms of activities rather than objects and operations supported by UCON. 

This comparison overview between ACAC and other related models strengthens the fact that how our proposed ACAC model distinctly supports `active' decision control and enforcement considering dynamic situations and scalability in distributed IoT-based smart systems with thousands of connected devices performing multiple activities in a dynamic environment.



\vspace{-2mm}
\section{Towards ACAC Security Models}
\label{Formal_ACAC}
\begin{figure}[!t]
    \centering
    \includegraphics[width = 0.6\textwidth]{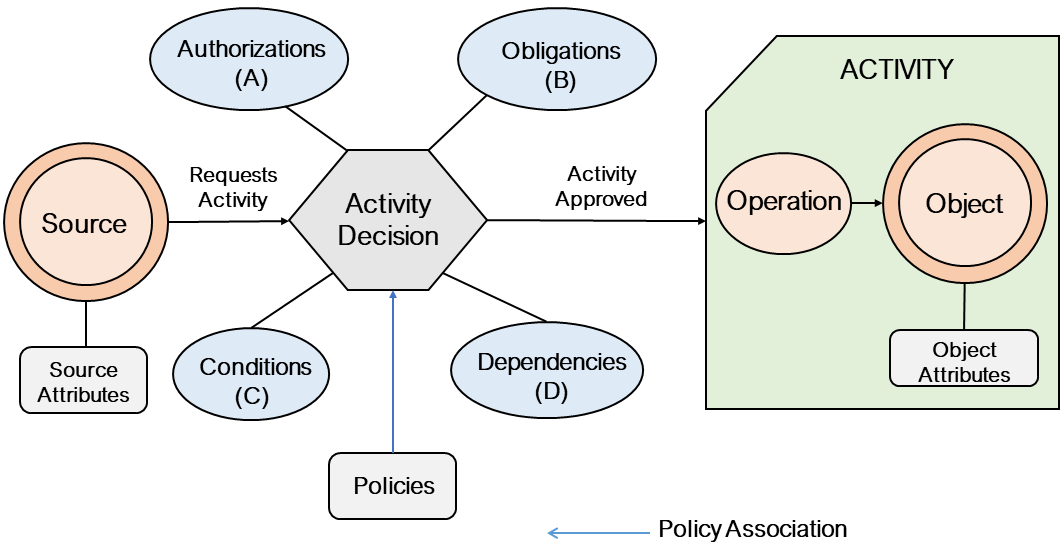}
    \centering
    \caption{ACAC Model Components \cite{mawla2022bluesky}}
    \label{fig-model}
\end{figure}

%
An \textbf{activity} is a prolonged event that is initiated by a source and occurs on an object for a certain period of time. The authors in \cite{gupta2021towards, mawla2022bluesky} motivated and proposed the activity-centric access control (ACAC) model components as shown in Figure \ref{fig-model} and described as follows. A source (S) can be a device, sensor, user, or an event in the system that requests an activity. An activity (ACT) is a long continuous task occurring for a period of time. An object (O) is an entity that performs the activity, such as an IoT device. To start an activity, a source will perform an operation (OP) on the object. When a source requests to initiate an activity, the decision depends on four components: authorizations (A), obligations (B), conditions (C), and activity dependencies (D) in the system. Authorizations define the right of a source to initiate an operation on an object. Source and object attributes take part in the authorizations. Obligations are the required tasks that must be fulfilled by the same requesting source or a different source in the system. Conditions are system or environmental factors related to satisfying the requested activity. Dependency on activities reflects relationships between single or multiple device activities in a system. 
 For example, in smart manufacturing, a robotic arm is requested to initiate \textit{painting} a box. If the robotic arm is currently \textit{washing} the product, it cannot be allowed immediately to paint the box. Here \textit{painting} and \textit{washing} are dependent activities. Our ultimate goal is to build an \textbf{active} security model for smart and collaborative systems utilizing all these components. However, with evolving different business needs and complexities, system designers and security administrators should be flexible in implementing some or all of these factors.
Accordingly, we define a family of four basic ACAC sub-models as $\mathrm{ACAC_{A}}$,  $\mathrm{ACAC_{B}}$, $\mathrm{ACAC_{C}}$, and $\mathrm{ACAC_{D}}$ for the proposed consolidated ACAC model, referred as $\mathrm{ACAC_{ABCD}}$. Each one of $\mathrm{ACAC_{A}}$,  $\mathrm{ACAC_{B}}$, $\mathrm{ACAC_{C}}$, and $\mathrm{ACAC_{D}}$ is a family of models. $\mathrm{ACAC_{A}}$ defines a family of models that define the authorization factor in a variety of ways to accommodate different application requirements. It considers the authorization factor only when deciding on an activity. $\mathrm{ACAC_{B}}$ handles the obligations factor, $\mathrm{ACAC_{C}}$ considers the impact of system and environmental conditions on an activity. $\mathrm{ACAC_{D}}$ incorporates the dependencies between different activities in all stages of the life cycle of a requested activity by checking and updating the current states of the dependent activities. 

\begin{figure}[!t]
\centering
\includegraphics[width=0.57\columnwidth]{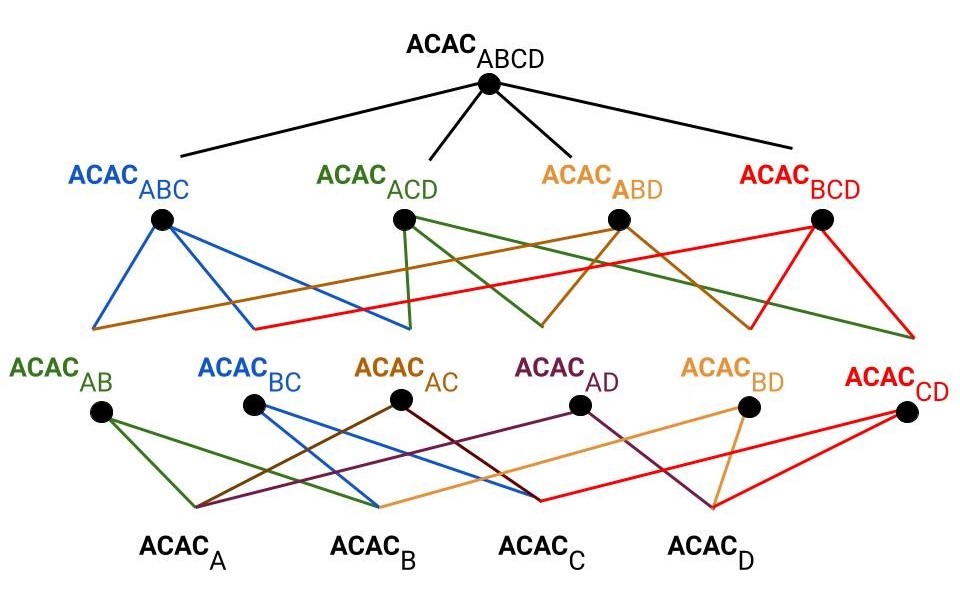}
\caption{Combinaton of $\mathrm{ACAC_{ABCD}}$ Core Models.}
\label{ACAC core models}
 \vspace{-3mm}
\end{figure}

Our proposed $\mathrm{ACAC_{ABCD}}$ model provides the \textbf{active} decision control by incorporating all of these decision factors  \cite{mawla2022bluesky}. Active decision control is defined as based on the real-time working environment considering authorizations, obligations, conditions, and dependencies on activities \cite{mawla2022bluesky}. Considering the complexity, in Figure \ref{ACAC core models}, we show how the combination of $\mathrm{ACAC_{ABCD}}$ core models are created from the basic models ($\mathrm{ACAC_{A}}$, $\mathrm{ACAC_{B}}$, $\mathrm{ACAC_{C}}$, and $\mathrm{ACAC_{D}}$).
We put the basic models at the bottom level, which includes individual models for each decision component (A-B-C-D). At the next two levels, models are composed of two and three models, respectively, from the immediate lower levels. As shown in Figure \ref{ACAC core models}, $\mathrm{ACAC_{ABCD}}$ is the final comprehensive model which combines the four sub-models. 
In order to consider the \textit{active} security needs, in this paper, our focus is to develop formal sub-models for the dependency (D) factor considering the relationship of activities, referred to as $\mathrm{ACAC_{D}}$ models. To our understanding and literature review, previous access control models have not considered these run-time dependencies as an active security factor, which is critical in smart connected and collaborative systems. The $\mathrm{ACAC_{D}}$ is mapped to $\mathrm{ACAC_{1}}$ model in the incrementally developed framework discussed by Mawla et al. \cite{mawla2022bluesky}. In our future work, we will develop the holistic $\mathrm{ACAC_{ABCD}}$ model considering the $\mathrm{ACAC_{A}}$, $\mathrm{ACAC_{B}}$,  $\mathrm{ACAC_{C}}$, and $\mathrm{ACAC_{D}}$ basic models.

\subsection{Mutability of Activities}
\begin{figure}[!t]
    \centering
   \includegraphics[width=.6\columnwidth]{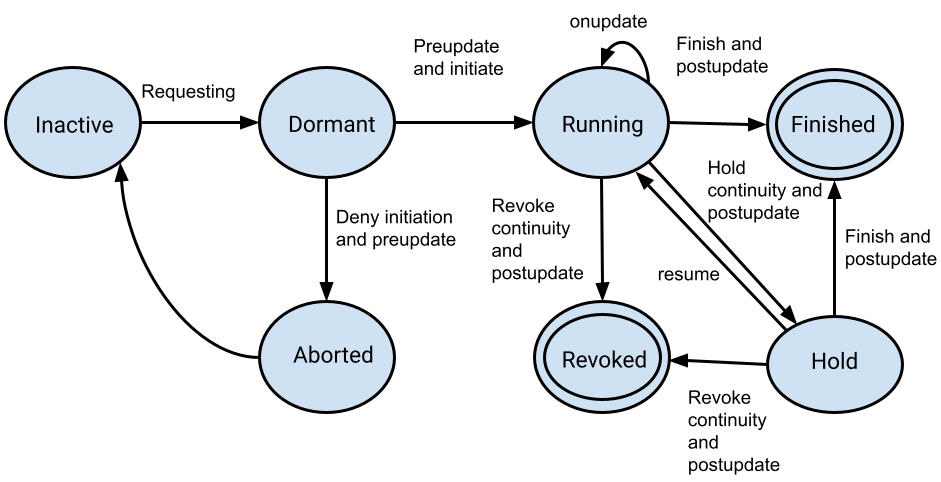}
    \centering
    \caption{State transition of an activity with required updates in an activity life-cycle.}
    \label{fig-transition}
     \vspace{-3mm}
\end{figure}

One of the ACAC model's unique characteristics is that the \textit{activities} in the system are mutable. \textbf{Mutable} activities can update their states (as discussed by Mawla et al. \cite{mawla2022bluesky}) as a consequence of the decision process of initiation, continuity, holding, completion, or revocation of an activity. In our models, \textbf{mutability} reflects the process of changing the state of mutable activities. In case of \textbf{immutable} activity, \textit{no outside} factor can change the activity state, and activity will complete its task while transitioning within its pre-defined course of states. 
Figure \ref{fig-transition} includes the states that an activity can have and shows the transitions between different states. An activity is in \textbf{inactive} state if it is not requested yet. When the activity is requested, the activity is in \textbf{dormant} state, and dependencies on other activities are assessed to see if the activity is allowed to be initiated. The dependent activities can be mutable and must change their states (if required) to allow the requested activity. In that case, the required pre-updates (updates before initiating an activity) on the dependent activities take place. Thus, the requested activity is invoked and goes to the \textbf{running} state. If the required pre-updates or any required condition cannot be fulfilled, the requested activity is denied and go to the \textbf{aborted} state. In the \textbf{running} state of activity, there can be required ongoing updates (updates during the execution of an activity to continue the execution) on the dependent activities. From the \textbf{running} state, an activity can be on \textbf{hold}, \textbf{finished}, or \textbf{revoked}. \textbf{Hold} state indicates a temporary suspension of the running activity due to any contextual conditions. Any required post update takes place after the activity goes to the \textbf{hold} state. From \textbf{hold} state the activity can resume and goes to the \textbf{running} state again. Otherwise, it can be \textbf{revoked} or \textbf{finished} based on the contextual conditions. The activity goes to a \textbf{revoked} state from the \textbf{running} state if the ongoing required updates (or ongoing conditions) are not fulfilled. \textbf{Finished} state indicates that the activity is completed and already served its purpose. Note that, from \textbf{finished} and \textbf{revoked} states, the requested activity goes back to the \textbf{inactive} state after the post-dependency check and update (if required). In Figure \ref{fig-transition}, the names of the states are more intuitive which helps in a better understanding of an activity's life-cycle than shown in \cite{mawla2022bluesky}. The transitions between activity states reflect the mutability of activities. It is a significant and distinguishable factor of ACAC compared with other access control models. 
In next subsection, we formally propose sub-models for $\mathrm{ACAC_{D}}$ which considers the mutability of activities. 

\begin{figure}[!htb]
    \centering
   \includegraphics[scale=0.5]{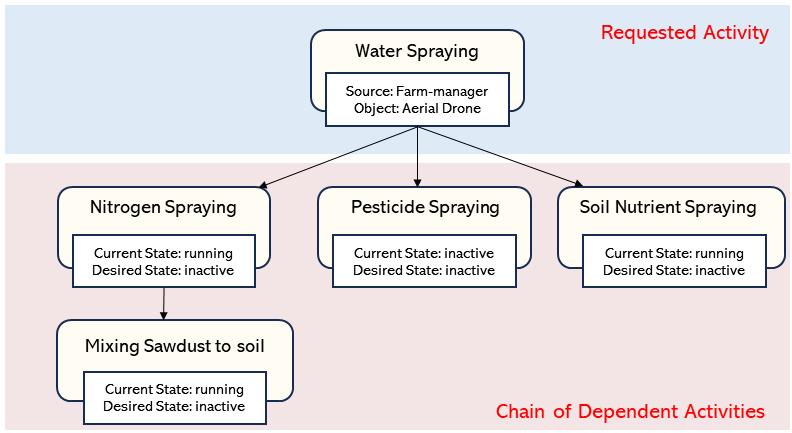}
    \centering
    \caption{Example of Chain of Dependencies.}
    \label{fig-chain_of_dependencies_example}
     \vspace{-3mm}
\end{figure}
\subsection{Chain of Dependencies}
\label{subsec:chain_of_dependencies}

A chain of dependencies refers to a series of dependencies where the dependency extends further down the line. In this paper, our goal is to inspect and analyze the dependent activities and control the mutability of these activities' states corresponding to a requested activity and its state transitions. In case the dependent activities, in-turn, they have some dependencies, i.e. "\textit{dependencies of dependencies}", we must ensure all the dependent activities are in their desired states before taking any decision on the requested activity. In such a scenario of a dependency chain, the system will wait to reach an independent activity (an activity that does not have any dependency) before any decision is made. We refer the requested activity as the "root" of the dependency chain and any dependent activity which depends on another activity for the state change is referred to as the "parent" of that dependent activity.

Figure \ref{fig-chain_of_dependencies_example} shows an example of a dependency chain corresponding to a requested activity "Water Spraying". This is requested by a Farm-manager and the system finds an "Aerial Drone" available to start "Water Spraying". Further, before allowing "Water Spraying" to start, we find three pre-dependent activities ("Nitrogen Spraying", "Pesticide Spraying" and "Soil Nutrient Spraying") which are shown in the first level of dependency in the colored portion of the chain of dependencies. The next level of dependent activities requires to be in the desired states according to the current and desired states of the previous dependent activities. For instance, in the figure, "Mixing Sawdust to soil" is a dependent activity according to the current and desired state of "Nitrogen Spraying". In such scenarios with "dependencies of dependencies", we only can update the state of activity when all dependent activities are in their desired states. This requires the system to find the chain of dependencies and update accordingly. In Section \ref{chain_of_dependencies}, we delve into the issue of the chain of dependencies. Throughout this section, we thoroughly examine the associated challenges and propose potential solutions to tackle this problem.

\subsection{$\mathrm{ACAC_{D}}$ Formal Models}
Dependencies on activities (D) are created due to the relationships among activities. The activities can be on the same or different devices. As characterized by Gupta and Sandhu \cite{gupta2021towards}, related activities can be characterized as ordered, concurrent, temporary, precedent, dependent, conditional, and incompatible. In this paper, we are not trying to develop a policy language for $\mathrm{ACAC_{ABCD}}$. Instead, we focus on formalizing the  $\mathrm{ACAC_{D}}$  models, which support the mutability of activities for active access control.
\begin{table}[!t]
\setlength{\tabcolsep}{4pt}
\renewcommand{\arraystretch}{1}
\centering
\caption{Family of $\mathrm{ACAC_{D}}$ sub-models}
\vspace{-4mm}
\scalebox{0.9}{%
\begin{tabular}{*{15}{|p{0.5cm}|p{2cm}|p{2cm}|p{1.5cm}|p{1.5cm}}}
\hline
 & \textbf{Immutable (0)} & \textbf{Pre-update (1)}& \textbf{Ongoing-update (2)} & \textbf{Post-update (3)} \\
\hline

preD&Y&Y&N&Y\\
\vspace{-2mm}
onD&Y&N&Y&Y
\\
\hline

\end{tabular}}
\label{model_space}
\vspace{-3mm}
\end{table}


Table \ref{model_space} shows the criteria for defining $\mathrm{ACAC_{D}}$ sub-models. The models are classified based on two parameters: (a) When the dependencies on related activities are checked to take any decision on the requested activity.  Decisions can be made \textit{pre} i.e., before allowing the requested activity to start (referred to as preD) or \textit{ongoing}, meaning while the requested activity is running (referred to as onD); 
(b) At which phase does the model support changing the states of dependent activities. The dependent activities can be either immutable or mutable, however, for immutable activities, model cannot update the states and may result in activity request denial. 
We denote the case as `0' when the current and the desire states of the dependent activities are checked without supporting the updates on dependent activities. On the other hand, if the model supports changing the states of dependent activities, then 
state updates are possible before (pre), during (ongoing), or after (post) the requested activity is performed. These cases are denoted as `1', `2', and `3', respectively. In all cases, the dependent activities can be both immutable and mutable, however, updates on dependent activities can be possible in `1', `2', and `3' for mutable activities.

In Table \ref{model_space}, cases marked as `Y' indicate the more practical scenarios considering when a decision is made, and  when dependent activities change state. Cases marked by `N' indicate that such scenarios are not practically useful. If the decision is taken before allowing the requested activity, updates on the dependent activities can occur before (pre) and after (post) the requested activity is performed. Without ongoing-decision, there is no need to have ongoing-update as a part of mutability, and is thus marked as `N'. For example, dependent activity B must be started before allowing requested activity A to start, and 
B should be revoked after A is finished. This case can be handled using pre and post update of B as a consequence of the initiation of A, and does not require ongoing-updates on B. 
However, if the decision is taken while the requested activity is ongoing, updates on the dependent activities can occur during (ongoing) and after (post) the requested activity is performed. In case of an ongoing-decision, the activity is already initiated. 
Thus, onD does not consider the pre-updates on the dependent activities and is marked as `N' for pre-update (1) of the onD case. The six `Y's in Table \ref{model_space} define six basic $\mathrm{ACAC_{D}}$ sub-models, which will be formalized in the following sections.

Different sub-model combinations of $\mathrm{ACAC_{D}}$ will be required for different type (pre, ongoing or post) of updates to solve all the recursive dependencies, as each sub-model defines a specific type of update. 
 Further, the dependent activities can be on the same or different device on which the activity is requested. Moreover, the dependent activity may be initiated by different objects in the system. In our model, the system chooses the object which can fulfil the activity, as will be discussed in the following sections.

\begin{figure}[!htb]
\centering
\includegraphics[scale=.4]{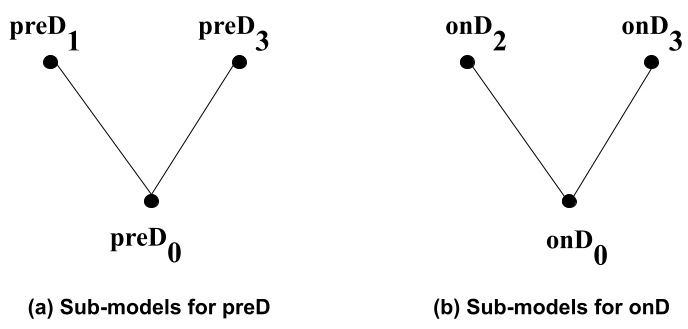}
\vspace{-4mm}
\caption{Categorization of $\mathrm{ACAC_{D}}$ sub-models.}
\label{ACAC_D_core_models}
\end{figure}
In Figure \ref{ACAC_D_core_models}, we show how the family of $\mathrm{ACAC_{D}}$ model is categorized into different sub-models. The `0' cases  for both preD and onD models only support checking the current and desired states of the activities, without any state updates. The `1', `2', and `3' cases supporting mutability add update procedures for the dependent mutable activities, and thus, inherit the basic components from the corresponding `0' cases. It should be noted that if the dependent activity is immutable, no state updates are allowed, and will result in activity request denial if the current and the desired states do not match.
We formally discuss the  components for each sub-model in the  following subsection.

In real-world use-cases, the activity-centric approach may need a \textbf{combination of two or more $\mathrm{ACAC_{D}}$ sub-models} checking pre-, ongoing, and post-dependencies. However, for clarity, we will formalize the behavior of the sub-models individually, and in our prototype implementation in Section \ref{sec:implementation}, we experiment with a more holistic multi-model comprehensive use case scenario.
\begin{table}[!htb]
\setlength{\tabcolsep}{6pt}
\renewcommand{\arraystretch}{1}
\caption{Introduction to basic sets and functions used in the definitions and algortihm}
\begin{tabular*}{\linewidth}{l}
\hline
\textbf{Basic Sets:} \\
- $S$, $O$, $OP$ and $ACT$ are finite sets of sources, objects, operations and activities in the system respectively.\\
- $ACT_R$ and $ACT_D$ is a set of requested and dependent activities such that $ACT_R$ = $ACT_D$ = $ACT$.\\
- $ACT_{DoD}$ is a set of dependent of dependent activities where $ACT_{DoD}$ = $ACT$\\
- $ST$ is a finite set of states of activities where $ST$ = \{inactive, dormant, aborted, running, hold, revoked, finished\}.\\
- $ST_{CR}$ and $ST_{DR}$ is finite sets of current and desired states of an activity where $ST_{CR}$ = $ST_{DR}$ = $ST$.\\
\textbf{Common Functions for Definitions:}\\
- $getObject: ACT_R \longrightarrow O$, mapping requested activity to an object. \\
 - $getOperation: ACT_R \times O \longrightarrow OP$, mapping a requested activity and an appropriate object to an operation to\\
 \;\; execute the activity.\\
 - $getDA: ACT_R \times O \longrightarrow 2^{ACT_D}$, mapping a requested activity and object to a set of dependent activities.\\
 \textbf{Common Functions for Definitions and Algorithms:}\\
 - $getCurrentSt: ACT \longrightarrow ST_{CR}$, mapping activity to its current state.\\

 
\textbf{Algorithm Functions:}\\



- $assignedDesiredSt:ACT_D \longrightarrow \{\emptyset ,ST_{DR}\}$, mapping a dependent activity to empty set or a desired state.\\

- $getBinarySemaphoreValue: ACT_D \longrightarrow$ \{0 , 1\}, 0 and 1 respectively indicate that the input dependent activity is \\
\;\;currently locked and unlocked.\\ 

- $hasConflictingDesiredSt: ACT_D
\longrightarrow$ \{TRUE , FALSE\}, TRUE indicates the input dependent activity has conflicting \\
\;\;(multiple) desired states and FALSE indicates it has no conflicting desired state.\\

 - $getDoDA: ACT_D \times ST_{CR} \times ST_{DR} \longrightarrow 2^{ACT_{DoD}}$, mapping a dependent activity, the current state of the dependent \\
 \;\;activity and a desired state of the dependent activity to a set of dependent of dependent activities.\\

 - $getDesiredDoDASt: ACT_D \times ST_{CR} \times ST_{DR} \times ACT_{DoD} \longrightarrow ST_{DR}$, mapping a dependent activity, the current state\\
 \;\;of dependent activity, desired state of dependent activity, and a dependent of dependent activity to a desired state.\\
 
\hline
\end{tabular*}
\label{set_and_function}
\end{table}
\setlength{\textfloatsep}{1pt}

\textbf{Table \ref{set_and_function}} elaborates the basic sets and functions we use in the formal definitions (1-6) and Algorithms (\ref{alg:algorithm1}, \ref{alg:algorithm2}, \ref{alg:algorithm3}). $S$, $O$, and $OP$ are the finite sets of sources, objects, and operations in the system [in Figure \ref{fig-model}, source is shown in a circle in the left part, and operation and object are shown respectively in elliptical and circle shape in the green part]. $ACT$ is a finite set of activities that can be performed in the system. $ACT_R$, $ACT_D$, $ACT_{DoD}$ are the finite sets of requested activities, dependent activities, and dependent of dependent activities respectively which are equivalent to the set of activities, $ACT$, formally we can say $ACT_R$ = $ACT_D$ = $ACT_{DoD}$ = $ACT$. $ST$ is the finite set of the activity states which is defined in the system as $\{inactive, dormant, aborted, running, hold, revoked, finished\}$. $ST_{CR}$ and $ST_{DR}$ are the finite set of current and desired states which are equivalent to the set, $ST$, formally we can say $ST_{CR}$ = $ST_{DR}$ = $ST$. The function $getObject$ maps a requested activity to the most suitable object to perform the activity in the system. This function can be called using a requested activity $act \in ACT$ and provides the most suitable object $o \in O$. $getOperation$ function determines the corresponding operation to start the requested activity
on the chosen object by the system. $getDA$ function maps a requested activity and its corresponding object to a set of dependent activities ($ACT_D$). The dependent activities for a particular requested activity can vary depending on the corresponding object. $getCurrentSt$ function maps an activity to a current state. $assignedDesiredSt$ function maps a dependent activity to an empty set or a desired state. This function is used to store a currently assigned desired state for a dependent activity. $getBinarySemaphoreValue$ function is used to provide the currently assigned value (0 or 1) for a dependent activity meaning that this activity is locked (cannot change the state) or unlocked by another activity. $hasConflictingDesiredSt$ function maps a dependent activity to TRUE or FALSE meaning whether that dependent activity has conflicting (multiple) desired states or not. $getDoDA$ function takes the input of a dependent activity, the current and desired state of this activity, and provides a set of activities which we call dependent of dependent activities. We refer `$DoD$' subscript to "dependent of dependent". To get the desired state of a dependent of dependent activity, we use the function $getDesiredDoDASt$ which maps a dependent activity, its current and desired state and a dependent of this dependent activity to a desired state. Apart from the basic sets and function in Table \ref{set_and_function}, we use two more functions from the algorithms (elaborated in Section \ref{chain_of_dependencies}) in the model definitions. One is \textbf{RECURSIVE-CHECK-OF-DEPENDENCIES-WITH-CONFLICT-DETECTION($da, da\_current\_st, da\_desired\_st$)}, which is a function in Algorithm \ref{alg:algorithm1} that recursively checks if an activity, $da$ has dependencies to transition from $da\_current\_st$ to $da\_desired\_st$ and for each activity, it detects whether the activity has conflicting desired states (multiple desired states) or not and stores the information. Another function is \textbf{RECURSIVE-UPDATE}(\textit{da, da\_current\_st, da\_desired\_st}) function from Algorithm \ref{alg:algorithm2} which recursively handles the state check and update process for all dependencies (including chain of dependencies) of a dependent activity, $da$.


\subsubsection{\textbf{$\mathrm{ACAC_{preD}}$ - pre-Dependency models}}

$\mathrm{ACAC_{preD}}$ models utilize the dependencies related to the decision process before the initiation of the requested activity. $\mathrm{ACAC_{preD}}$ has three sub-models (stated in Figure \ref{ACAC_D_core_models} (a) $\mathrm{ACAC_{preD_0}}$ model checks the pre-dependencies that are required to allow the requested activity. 
$\mathrm{ACAC_{preD_0}}$ model does not support mutability (i.e. cannot update dependent activity states). $\mathrm{ACAC_{preD_1}}$ model allows pre-updates on the dependent activities that require to be in specific states to allow the requested activity. $\mathrm{ACAC_{preD}}$ does not have ongoing-update model since ongoing-update without ongoing-decision does not need to be considered as a part of mutability. 
 Post-updates on dependent activities as a consequence of the pre-decision process are handled in $\mathrm{ACAC_{preD_3}}$ model. The following three definitions formalize $\mathrm{ACAC_{preD}}$ models. We elaborate the basic sets and functions in Table \ref{set_and_function} and use the necessary sets and functions in these definitions from the table.\\




\textbf{Definition 1. 
$\mathrm{ACAC_{preD_0}}$: Pre-dependency checking model for pre-dependent activities.}
$\mathrm{ACAC_{preD_0}}$ model checks the current and desired states of the pre-dependent activities before allowing a requested activity.
This model does not have any update procedure for state change and cannot support mutability of dependent activities.
$\mathrm{ACAC_{preD_0}}$ consists of the following components (shown in Figure \ref{fig-model}), and explained later:

\noindent


-- $S, O, OP, ACT, ACT_R, ACT_D$, $ST$, $ST_{CR}$, $ST_{DR}$ are finite sets of sources, objects, operations, activities, requested activities, dependent activities, activities' states, current states and desired states respectively [elaborated in Table \ref{set_and_function}]. A source $s \in S$ requests to perform an activity $act \in ACT$, defined as $request (s, act)$. To satisfy this activity request (formally stated as, $request (s, act)$ = $True$), the system will first specify an appropriate object $o \in O$, and perform an operation $op \in OP$ (Note that, whether source $s$ is allowed to perform an operation $op$ on an object $o$ is determined by the authorization model $\mathrm{ACAC_{A}}$). Then, the system will check activity dependencies based on the corresponding to the the requested activity and the object, using $getDA$ function.\\
 -- $getDesiredPreDASt: ACT_R \times ACT_D \longrightarrow ST_{DR}$\\\hspace*{0.5in}\Comment{\texttt{\textit{\textbf{[mapping a requested activity, and a dependent activity to a desired state.]}}}}\\
-- $preD(act : ACT, o : O) \longrightarrow \{True, False\}$,\\
\hspace*{1in}defined as
$ \bigwedge_{(da\in getDA(act,o)} getCurrentSt(da)$ = $getDesiredPreDASt(act, da)$. \\
-- $allowed(s : S, o : O, op : OP, act : ACT) \Rightarrow preD(act, o)$\\
\noindent
$\mathrm{ACAC_{preD_0}}$ model consists of sources ($S$), objects ($O$), operations ($OP$), activities ($ACT$), requested activities ($ACT_R$), dependent activities ($ACT_D$), finite set of activities' states ($ST$), activities' current states ($ST_{CR}$) and activities' desired states ($ST_{DR}$). The function $getObject$ maps a requested activity to the most suitable object $o \in O$ to perform the activity in the system. $getOperation$ determines the corresponding operation to start an activity on the chosen object, o. More than one combination of activity and object can be mapped to an operation. The function $getDA$ computes the set of dependent activities, decided based on the activity $act \in ACT$ and the corresponding object $o \in O$. Note that the dependencies are dynamic, and can change based on conditions (C) and contextual factors. This is a many-to-one mapping function where each combination of activity and object can be mapped to a set of activities. The function $getCurrentSt$ is used to get the current state of an activity and $getDesiredPreDASt$ is used to determine the desired states of pre-dependent activities (activities that need to be checked before starting activity $act$). $getCurrentSt$ and $getDesiredPreDASt$ are many-to one mapping functions. 


$preD$ is a functional predicate that takes the requested activity and the corresponding object (since dependencies can change based on which object is performing the activity) as inputs, and return $True$ or $False$ by comparing the current and desired states of all pre-dependent activities. $True$ indicates that all dependent activities' current states are in the desired states.
$False$ indicates that at least one dependent activity is not in the desired state to allow the requested activity to be initiated. 
To allow the request, formally stated as $request (s, act)$ = $True$, the $allowed (s, o, op, act)$ function (which decides $s$ can perform operation $op$ to start the activity $act$ on the object $o$) should evaluate to $True$. The $allowed$ function returns $True$ if $preD$ evaluates to $True$. Note that, we use the $implies$ $(\implies)$ connective where the right hand side of the connective is necessary but not sufficient since authorization (A), oBligations (B) and conditions (C) also be checked for the left hand side to be $True$. There is no update procedure in this model.\\\\
\textbf{Example 1.} In smart manufacturing, a \textit{$robot$} is trying to make a \textit{$forceGeneration$} activity request, stated as \\
$request(robot, forceGeneration)$.\\
-- $S = \{robot\}$\\
-- $O = \{motor\}$\\
-- $OP = \{turnOn, turnOff\}$\\
-- $ACT = \{forceGeneration, vibrationMonitoring\}$\\
-- $ST = \{inactive, dormant, aborted, running, hold, revoked, finished\}$\\
-- $getObject(forceGeneration) = motor$\\
-- $getOperation(forceGeneration, motor) = turnOn$\\
-- $getDA(forceGeneration, motor) =\{vibrationMonitoring\}$\\
-- $getCurrentSt(vibrationMonitoring) = running$\\
-- $getDesiredPreDASt(forceGeneration, vibrationMonitoring) = running$\\
-- $preD(forceGeneration, motor)$ = $True$;\\
-- $allowed(robot, motor, turnOn, forceGeneration) \Rightarrow$ $preD(forceGeneration, motor)$
\sloppypar
In this example, to satisfy the request made by the source $robot$, we get the corresponding object $motor$ and operation $turnOn$ for the requested activity.  
The set of dependent activities for $forceGeneration$ consists of $vibrationMonitoring$. The desired state of $vibrationMonitoring$ is $running$. In this instance, the current state is same as the desired state for the only dependent activity. Thus, $preD(forceGeneration, motor)$ is $True$ as the necessary condition (comparing the current and desired states of the dependent activity) in
$preD(forceGeneration, motor)$ is fulfilled. The $allowed$ function also returns $True$ which decides that source $robot$ is allowed to perform the operation, $turnOn$ on the object $motor$ to initiate the requested activity, $forceGeneration$. \\

\textbf{Definition 2. $\mathrm{ACAC_{preD_1}}$: Pre-update model for pre-dependent activities.}
$\mathrm{ACAC_{preD_1}}$ model adds state update procedure for the pre-dependent activities (dependent activities that are required to be in desired state before initiation of the requested activity). These pre-dependent activities may, in-turn, be dependent on other activities.  
For example, starting the requested activity A depends on starting the dependent activity B. Activity B can't start until activity C has already started. In such situations, we have to update the states of the pre-dependent activities in a recursive way, where we explore the "\textit{dependencies of dependencies}" until we find a dependent activity that does not have any dependent activity before changing its state or all dependent activities need to be already in their desired states. 
Algorithm \ref{alg:algorithm1} includes a function named RECURSIVE-CHECK-OF-DEPENDENCIES-WITH-CONFLICT-DETECTION where a dependent activity, the current and desired state of that activity are passed as parameters. We check if this dependent activity has any conflicting (multiple) desired states or not and store this information. Note that, this function is recursive and we recursively detect the conflicting desired states for all "dependencies of dependencies" along with the dependent activity (explained in Section \ref{chain_of_dependencies}). In Algorithm \ref{alg:algorithm2} in Section \ref{chain_of_dependencies}, we have a function named RECURSIVE-UPDATE. In this function, we pass the parameters for a dependent activity, its current state and a desired state of this dependent activity. This function returns the desired state after checking and updating (if necessary) all the "dependencies of dependencies". We explain Algorithm \ref{alg:algorithm2} in Section \ref{chain_of_dependencies} describing the way it works with the recursive update procedure of "chain of dependencies".
Conceptually, $\mathrm{ACAC_{preD_1}}$ model is an extension to $\mathrm{ACAC_{preD_0}}$ as it adds the pre-update procedure when \textit{allowed} function returns False. Thus, to satisfy the activity request $request (s: S, act: ACT) = True$, $\mathrm{ACAC_{preD_1}}$ model allows updating the states of the pre-dependent activities using the following $preUpdate(act)$ function defined as.

\noindent
-- $preUpdate(act, o)$: \hspace*{1.1in}\Comment{\texttt{\textit{\textbf{[Function Definition]}}}}\\
\hspace*{.12in}$(\forall da \in getDA(act, o)).$\\
\hspace*{.25in}[RECURSIVE-CHECK-OF-DEPENDENCIES-WITH-CONFLICT-DETECTION($da, getCurrentSt(da),\\ \hspace*{.4in}getDesiredPreDASt(act, da)$) \\
\hspace*{.25in}$getCurrentSt(da) \neq getDesiredPreDASt(act, da) \\
\hspace*{.5in}\Rightarrow getCurrentSt(da) =$ RECURSIVE-UPDATE$(da, getCurrentSt(da), getDesiredPreDASt(act, da))]$\\
-- $preUpdate(act, o) \Rightarrow allowed(s, o, op, act) == False$\hspace*{1.4in}\Comment{\texttt{\textit{\textbf{[Function Call]}}}}

$\mathrm{ACAC_{preD_1}}$ model introduces the $preUpdate$ function to update the states of the pre-dependent activities that are required to be in specific states for the initiation of the requested activity $act$ on the object $o$. In this function, we iterate a loop for all the dependent activities where the current state of each dependent activity is updated to the desired state if it is not in the desired state at the time of the request. Before updating the current state of each dependent activity, we check whether the dependent activity (including its dependencies) in the loop has conflicting desired states or not utilizing the function, RECURSIVE-CHECK-OF-DEPENDENCIES-WITH-CONFLICT-DETECTION in Algorithm \ref{alg:algorithm1}. After that, we call the RECURSIVE-UPDATE function in Algorithm \ref{alg:algorithm2} by the dependent activity, its current state, and the desired state and resolve the state-updates for "chain of dependencies" where it is required. This function returns the desired state and we update the current state to the desired state. $preUpdate$ function is called when the $allowed$ function returns $False$ as the current states of all the dependent activities are not in their desired states. 
For simplicity, issues like who will update the state of the activity
and underlying technical implementation of the update procedure is left unspecified in this paper.\\

\textbf{Example 2.} In smart home, the $houseOwner$ is trying to make the request for the activity, $playingNews$. The request is stated as $request(houseOwner, playingNews)$.\\
-- $S = \{houseOwner\}$\\
-- $O = \{TV, googleHome\}$\\
-- $OP = \{turnOn, turnOff\}$\\
-- $ACT = \{playingSong, playingNews\}$\\
-- $ST = \{inactive, dormant, aborted, running, hold, revoked, finished\}$\\
-- $getObject(playingNews) = TV$\\
-- $getOperation(playingNews, TV) = turnOn$\\
-- $getDA(playingNews, TV)
= playingSong$\\
-- $getCurrentSt(playingSong) = running$\\
-- $getDesiredPreDASt(playingNews, playingSong) = inactive$\\
-- $preD(playingNews, TV)= False$\\
-- $allowed(houseOwner, TV, turnOn, playingNews) \Rightarrow preD(playingNews, TV)$\\
-- $preUpdate(playingNews) \Rightarrow preD(playingNews, TV) == False$



In Example 2, to satisfy $request(houseOwner, playingNews)$, we get the corresponding object $TV$ and the operation $turnOn$. The set of dependent activities (provided by $getDA(playingNews, TV)$) for $playingNews$ consists of $playingSong$. In this instance, the current state of $playingSong$ is \textit{running}, which is not the same as the desired state \textit{inactive}. Thus, $preD$ is false, and so is the $allowed$ function. Therefore, the model updates the current state of $playingSong$ to \textit{inactive} using the $preUpdate(playingNews)$ function. Once updated, the request $request(houseOwner, playingNews)$ is allowed.\\

\textbf{Definition 3. $\mathrm{ACAC_{preD_3}}$: Post-update model for dependent activities with pre-check.}
$\mathrm{ACAC_{preD_3}}$ model adds the post-update procedure which updates the states of the dependent activities after the requested activity is finished, revoked or on hold. Updating the states of these dependent activities accumulate the consequence of the requested activity. In pre-check, we check the pre-dependent activities that need to change their states after the completion or revocation of the requested activity. For example, a dependent activity B have already started to help executing the requested activity A. 
After A is finished, activity B is no longer needed. Thus, we make sure there are no unnecessary activities going on after the purpose is completed. In such cases, combination of pre-update and post-update models is more appropriate. However, we consider post-update as a separate procedure.
Conceptually, $\mathrm{ACAC_{preD_3}}$ model is an extension to $\mathrm{ACAC_{preD_0}}$ which adds the post-update procedure.

 \noindent
 -- $getDesiredPostDASt: ACT_R \times ACT_D \longrightarrow ST_{DR}$\\\hspace*{0.5in}\Comment{\texttt{\textit{\textbf{[mapping a requested activity which has either been on `hold', `finished' or `revoked', and a post-dependent activity to a desired state]}}}}\\
 -- $postD(act : ACT, o : O) \longrightarrow \{True, False\}$, defined as \\
\hspace*{.5in}$\bigwedge_{(da\in getDA(act,o))} getCurrentSt(da) = getDesiredPostDASt(act, da)$\\
--$postUpdate(act, o)$: \hspace*{1.1in}\Comment{\texttt{\textit{\textbf{[Function Definition]}}}}\\
\hspace*{.12in}$(\forall da \in getDA(act, o)).$\\
\hspace*{.25in}[RECURSIVE-CHECK-OF-DEPENDENCIES-WITH-CONFLICT-DETECTION($da, getCurrentSt(da),\\ \hspace*{.4in}getDesiredPostDASt(act, da)$) \\
\hspace*{.25in}$getCurrentSt(da) \neq getDesiredPostDASt(act, da) \\
\hspace*{.5in}\Rightarrow getCurrentSt(da) =$ RECURSIVE-UPDATE$(da, getCurrentSt(da), getDesiredPostDASt(act, da))]$\\
--$postUpdate(act, o)\Rightarrow postD(act, o) == False$\hspace*{1.4in}\Comment{\texttt{\textit{\textbf{[Function call]}}}}










$\mathrm{ACAC_{preD_3}}$ model includes the $postUpdate$ function to update the states of the dependent activities after the requested activity $act$ is performed. The $getDesiredPostDASt$ is a many-to-one function to get the desired states of the post-dependent activities. It maps the requested activity and a dependent activity to a desired state. 
Then the $postD$ function is evaluated checking the current and desired states of the post-dependent activities. In $postUpdate$ function, conflicting desired states are checked for all the post-dependent activities calling the RECURSIVE-CHECK-OF-DEPENDENCIES-WITH-CONFLICT-DETECTION function from Algorithm \ref{alg:algorithm1} followed by updating their current states to their corresponding desired states utilizing the RECURSIVE-UPDATE function from Algorithm \ref{alg:algorithm2}. 
This $postUpdate$ function is called when $postD$ returns $False$ (which means that the current states of all dependent activities are not in their desired states).\\

\textbf{Example 3.} In smart industry a, a $productionWorker$ is requesting $hydrotreating$ activity, 
formally stated as $request(productionWorker, hydrotreating)$.\\
-- $S = \{productionWorker\}$\\
-- $O = \{tankPump, hydrotreater\}$\\
-- $OP = \{turnOn\}$\\
-- $ACT = \{oilPumping, hydrotreating\}$\\
-- $ST = \{inactive, dormant, aborted, running, hold, revoked, finished\}$\\
-- $getOperation(hydrotreating, hydrotreater) = turnOn$\\
-- $getDA(hydrotreating, hydrotreater) = \{oilPumping\}$\\
-- $getCurrentSt(oilPumping) = inactive$\\
-- $getDesiredPostDASt(hydrotreating, oilPumping) = running$\\
-- $postD(hydrotreating, hydrotreater)= False$\\
 -- $postUpdate(hydrotreating)\Rightarrow postD(hydrotreating, hydrotreater) == False$\\
In Example 3, the requested activity is $hydrotreating$. This request was allowed and  has just finished. Now, we need to update the post-dependent activities of $hydrotreating$. We get the set of dependent activities for $hydrotreating$ (using $getDA(hydrotreating, hydrotreater)$ function) which consists of one activity, $oilPumping$ (assuming $oilPumping$ already served its purpose of activating $hydrotreating$). The current and desired states of $oilPumping$ are not same in this instance. Thus, the $postD$ function returns $False$. We call $postUpdate(hydrotreating)$ function where the current state of $oilPumping$ is updated to the desired state.
\vspace{-2mm}
\subsubsection{\textbf{$\mathrm{ACAC_{onD}}$ - Ongoing-Dependency Models}}
$\mathrm{ACAC_{onD}}$ models consider the dependencies on activities while the requested activity is ongoing. 
The ongoing decisions can be \textit{continue}, \textit{hold} or, \textit{revoke} the requested activity, and can impact dependent activities. Execution of the requested activity can be \textit{continued} if the ongoing dependent activities are in the desired states. If the dependent activities are mutable, their current states can be updated for the continuity of the requested activity. Otherwise, the execution of the requested activity will be \textit{revoked}. Besides that, holding the requested activity can accumulate any emergence or contextual situations. 
$\mathrm{ACAC_{onD}}$ has three sub-models (stated in Figure \ref{ACAC_D_core_models} (b)) based on if states of dependent activities can be updated and which phase the updates can occur as shown in Table \ref{model_space}. $\mathrm{ACAC_{onD_0}}$ model checks the current and desired states of the ongoing dependent activities.
$\mathrm{ACAC_{onD_0}}$ model does not support mutability. $\mathrm{ACAC_{onD_2}}$ allows updates on the states of the ongoing dependent activities as a consequence of the ongoing-decisions. $\mathrm{ACAC_{onD_3}}$ model checks and updates the post-dependent activity states that are related to the ongoing activity and decisions. $\mathrm{ACAC_{onD}}$ does not have the $\mathrm{ACAC_{onD_1}}$ model since the requested activity is already allowed and there is no reason to consider the pre-updates after allowing the activity.  
Since the ongoing dependent activities are checked during the execution of the requested activity, how frequently the dependencies are checked is unspecified, and left for the implementation details.\\


\textbf{Definition 4. $\mathrm{ACAC_{onD_0}}$: Ongoing-dependency checking model for ongoing dependent activities}\\
$\mathrm{ACAC_{onD_0}}$ model checks the dependencies on 
activities while the requested activity is running to decide \textit{continuity} or \textit{revocation} of the ongoing activity. 
There is no update procedure in this model. We need this model only to check if all the ongoing dependent activities are in their desired states or not. The model consists of the following components:

\noindent
A source $s \in S$ requests to perform an activity $act \in ACT$, defined as $request (s, act)$. Since, $\mathrm{ACAC_{onD_0}}$ model checks the ongoing dependencies on activities, the requested activity is assumed to be initially allowed.

\noindent
-- $allowed(s:S, o:O, op: OP, act:ACT) \Rightarrow True$\\
-- $getDesiredOnDASt: ACT_R \times ACT_D \longrightarrow ST_{DR}$\\\hspace*{0.5in}\Comment{\texttt{\textit{\textbf{[mapping a requested `running' activity, and an ongoing-dependent activity to a desired state.]}}}}\\
-- $onD(act:ACT, o:O) \longrightarrow \{True, False\}$, defined as \\
\hspace*{.2in}$\bigwedge_{(da\in getDA(act, o))} getCurrentSt(da) = getDesiredOnDASt(act, da)$.\\
-- $stopped(act:ACT, o:O) \Rightarrow onD(act, o) == False$
\\ 
\noindent
$\mathrm{ACAC_{onD_0}}$ model consists of sources ($S$), objects ($O$), operations ($OP$), activities ($ACT$), requested activities ($ACT_R$), dependent activities ($ACT_D$), finite set of activities' states ($ST$), activities' current states ($ST_{CR}$) and activities' desired states ($ST_{DR}$) [explained in Table \ref{set_and_function}]. 
$getObject$ function provides the corresponding object the activity is running on. $getOperation$ function provides the operation $op$ that is performed on object $o$ to initiate the requested activity, $act$. The $allowed$ function is $True$ since the requested activity is already assumed to be running currently, and the check is only made for ongoing decision. 
$getDA$ function computes the set of dependent activities for the ongoing activity, $act \in ACT$.
$getDesiredOnDASt$ is used to get the desired states of the ongoing-dependent activities. This function maps the requested `running' activity and a dependent activity to a desired state. 
$onD$ is a functional predicate which takes input of the requested activity and corresponding object (since dependencies can change based on the object which is performing the activity), and compares the current and desired states of all ongoing-dependent activities (and returns $True$ or $False$) to make a decision.  Ongoing dependencies are checked throughout the execution of the activity $act$ using the $onD$ function. If $onD$ returns $False$, the activity will be revoked which is handled using the $stopped$ function. We do not have any update procedure in this model.\\









\textbf{Example 4.} In smart farming, activity $cooling$ is requested by the $farmManager$ (formally stated as  $request (farmManager, cooling)$) and is assumed to be allowed. In the ongoing check, our model ensures the corresponding dependencies are fulfilled. \\
-- $S = \{farmManager\}$\\
-- $O = \{cooler, aerialDrone\}$\\
-- $OP = \{turnOff, turnOn\}$\\
-- $ACT = \{thermalImaging, cooling\}$\\
-- $ST = \{inactive, dormant, aborted, running, hold, revoked, finished\}$\\
-- $getObject(cooling) = cooler$\\
-- $getOperation(cooling, cooler) = turnOn$\\
-- $getDA(cooling, cooler) = \{thermalImaging\}$\\
-- $getCurrentSt(thermalImaging) = inactive$\\
-- $getDesiredOnDASt(cooling, thermalImaging) = running$\\
-- $onD(cooling, cooler) = False$\\
-- $stopped(cooling, cooler)$\\
In this example, $thermalImaging$ is an immutable and ongoing-dependent activity for $cooling$ to obtain the current temperature and relevant status of the environment. The desired state of $thermalImaging$ is $running$ to continue $cooling$. As the current state of $thermalImaging$ is $inactive$ (and cannot be changed) which is different from the desired state, $cooling$ will be revoked.\\
\textbf{Definition 5. $\mathrm{ACAC_{onD_2}}$: Ongoing-update model for ongoing dependent activities}\\
$\mathrm{ACAC_{onD_2}}$ model adds the update procedure to change the states (if not in desired state) of the ongoing dependent activities of a requested activity. The updates are required to allow the requested activity to \textit{continue}. 
For example, A is the requested activity which is executing and B is the dependent activity that should be $running$ to continue activity A. In this model, we can update the state of activity B from $inactive$ to $running$ to allow the activity A to \textit{continue}. $\mathrm{ACAC_{onD_2}}$ model includes a function $onUpdate$ for such ongoing updates. This model is an extension to $\mathrm{ACAC_{onD_0}}$ adding the ongoing update procedure.

\noindent
--$onUpdate(act, o)$: \hspace*{1.1in}\Comment{\texttt{\textit{\textbf{[Function Definition]}}}}\\
\hspace*{.12in}$(\forall da \in getDA(act, o)).$\\
\hspace*{.25in}[RECURSIVE-CHECK-OF-DEPENDENCIES-WITH-CONFLICT-DETECTION($da, getCurrentSt(da),\\ \hspace*{.4in}getDesiredOnDASt(act, da)$) \\
\hspace*{.25in}$getCurrentSt(da) \neq getDesiredOnDASt(act, da) \\
\hspace*{.5in}\Rightarrow getCurrentSt(da) =$ RECURSIVE-UPDATE$(da, getCurrentSt(da), getDesiredOnDASt(act, da))]$\\
--$onUpdate(act, o)\Rightarrow onD(act, o) == False$
\hspace*{1.4in}\Comment{\texttt{\textit{\textbf{[Function Call]}}}}


For the requested activity to \textit{continue}, ongoing-dependent activities may require state change.
 In $onUpdate(a)$ function, we iterate a loop for each ongoing dependent activity, check if the dependent activities and the dependent of dependent activities have conflicting (multiple) desired states or not (calling RECURSIVE-CHECK-OF-DEPENDENCIES-WITH-CONFLICT-DETECTION function from Algorithm \ref{alg:algorithm1}) followed by updating their current states by calling the RECURSIVE-UPDATE function in Algorithm \ref{alg:algorithm2} (with checking and updating the states of "chain of dependencies"). This $onUpdate$ function is called when $onD$ returns $False$ suggesting that not every dependent activity is in desired state.\\ 




\textbf{Example 5.} In smart farming, an ongoing activity is $cooling$ the greenhouse requested by the source $farmManager$ (formally stated as  $request (farmManager, cooling)$).\\
-- $S = \{farmManager\}$\\
-- $O = \{airCooler, humidifier\}$\\
-- $OP = \{turnOn, turnOff\}$\\
-- $ACT = \{cooling, humidifying\}$\\
-- $ST = \{inactive, dormant, aborted, running, hold, revoked, finished\}$\\
-- $getObject(cooling) = airCooler$\\
-- $getOperation(cooling, cooler) = turnOn$\\
-- $getDA(cooling, airCooler) = \{humidifying\}$\\
-- $getCurrentSt(humidifying) = inactive$\\
-- $getDesiredOnDASt(cooling, humidifying) = running$\\
-- $onUpdate(cooling)\Rightarrow onD(cooling, airCooler) == False$\\
In example 5, the ongoing activity is $cooling$ the environment of a greenhouse using the object $airCooler$. While $cooling$, if the humidity is low the $humidifier$ should be $running$ to continue $cooling$. In that case, $humidifying$ is an ongoing dependent activity for $cooling$. We call the $onUpdate(cooling)$ function and update the current state of $humidifying$ from \textit{inactive} to the  $running$ state as the $onD$ function returns $False$. This will ensure that the \textit{cooling} continues while $humidifying$ is running.\\ 



\textbf{Definition 6. $\mathrm{ACAC_{onD_3}}$: 
Post-update model for dependent activities with ongoing-check}

$\mathrm{ACAC_{onD_3}}$ model adds the update procedure for the dependent activities which may need state change when the requested activity is finished, on hold, or revoked, requiring ongoing check.
For instance, A is a requested activity and B is a dependent activity which needs to be started while A is running. After A is revoked, B should be stopped immediately. This is a post-update on B based on the decision taken on activity A while running (ongoing check). 
$\mathrm{ACAC_{onD_3}}$  model is an extension to $\mathrm{ACAC_{onD_0}}$ adding the post-update procedures.\\
\noindent
-- $getDesiredPostDASt: ACT_R \times ACT_D \longrightarrow ST_{DR}$\\\hspace*{0.5in}\Comment{\texttt{\textit{\textbf{[mapping a requested activity which has been `finished', `revoked', or on `hold', and a post-dependent activity to a desired state.]}}}}\\
 -- $postD(act : ACT, o : O) = \bigwedge_{da\in getDA(act, o)} getCurrentSt(da) = getDesiredPostDASt(act, da)$\\
--$postUpdate(act, o)$: \hspace*{1.1in}\Comment{\texttt{\textit{\textbf{[Function Definition]}}}}\\
\hspace*{.12in}$(\forall da \in getDA(act, o)).$\\
\hspace*{.25in}[RECURSIVE-CHECK-OF-DEPENDENCIES-WITH-CONFLICT-DETECTION($da, getCurrentSt(da),\\ \hspace*{.4in}getDesiredPostDASt(act, da)$) \\
\hspace*{.25in}$getCurrentSt(da) \neq getDesiredPostDASt(act, da) \\
\hspace*{.5in}\Rightarrow getCurrentSt(da) =$ RECURSIVE-UPDATE$(da, getCurrentSt(da), getDesiredPostDASt(act, da))]$\\
--$postUpdate(act, o)\Rightarrow postD(act, o) == False$
\hspace*{1.4in}\Comment{\texttt{\textit{\textbf{[Function Call]}}}}\\
 In this model, $getDesiredPostDASt$ function provides the desired state of a post-dependent activity. This function takes a requested activity and one of its dependent activity as input and returns a desired state for this dependent activity. $postD$ function checks the current and desired states of the post-dependent activities and returns $True$ or $False$ based on the outcome of the comparison between current and desired states of all post-dependent activities. In $postUpdate$ function, the current states of all the dependent activities are updated (to desired states) if they are not in the desired states (i.e., if $postD$ returns $False$). Before updating the states, we check if the dependent activity $da$ or any of its dependent activity has conflicting desired states or not. We call the RECURSIVE-CHECK-OF-DEPENDENCIES-WITH-CONFLICT-DETECTION function from Algorithm \ref{alg:algorithm1} (explained in Section \ref{chain_of_dependencies})  by passing a post-dependent activity, its current state and its desired state. After that, we call the RECURSIVE-UPDATE function in Algorithm \ref{alg:algorithm2} by passing a post-dependent activity, its current state, and its desired state and it returns the desired state after checking and updating the "chain of dependencies".\\

\textbf{Example 6.} In smart home, $floorCleaning$ was requested by the source $floorWorker$, stated as  $request (floorWorker, floorCleaning)$.\\
-- $S = \{floorWorker, sensor\}$\\
-- $O = \{vacuumCleaner, roboticArm\}$\\
-- $OP = \{turnOn, turnOff\}$\\
-- $ACT = \{movingObject, floorCleaning\}$\\
-- $ST = \{inactive, dormant, aborted, running, hold, revoked, finished\}$\\
-- $getObject(floorCleaning) = vacuumCleaner$\\
-- $getOperation(floorCleaning, vacuumCleaner) = turnOn$\\
-- $getDA(floorCleaning, vacuumCleaner) = \{movingObjects\}$\\
-- $getCurrentSt(movingObjects) = running$\\
-- $getDesiredPostDASt(floorCleaning, movingObjects) = inactive$\\
--$postUpdate(floorCleaning)\Rightarrow postD(floorCleaning, vacuumCleaner) == False$

In Example 6, we assume the activity $floorCleaning$ has been just finished which was running on the object, $vacuumCleaner$. For the continuity of this activity, $movingObjects$ by $roboticArm$ was running. The purpose of $movingObjects$ is done after $floorCleaning$ is finished. Thus, $movingObjects$ needs to be in $inactive$ state as a post-dependent activity. We update the state using the $postUpdate(floorCleaning)$ function.

\begin{figure}[!htb]
    \centering
\includegraphics[width=.5\columnwidth]{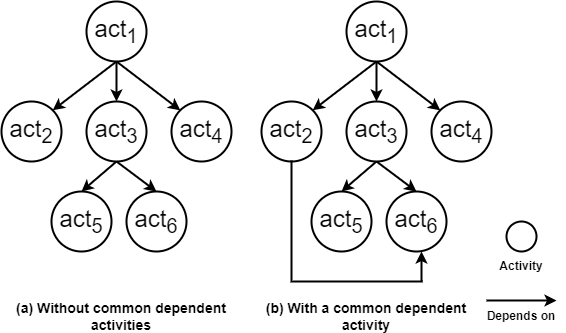}
    \centering
    \caption{Chain of Dependencies with Multiple Dependency Paths.}
    \label{fig-multiple_dependency_path}
\end{figure}
\vspace{-3mm}
\section{Challenges of Resolving Chain of Dependencies}
\label{chain_of_dependencies}

Chain of dependencies refers to "dependencies of dependencies" where one activity relies on another activity for the state transition, which in turn relies on some other activity and these sequence continues until there exists one independent activity which is not dependent on others for its state transition. In our proposed $\mathrm{ACAC_{D}}$ model, the dependencies are a set of activities that need to be in their desired states to allow the transition of a parent activity from its current state to the desired state.  We discuss the chain of dependencies with a relevant example in Section \ref{subsec:chain_of_dependencies}. There are few challenges when resolving the chain of dependencies that increase the complexity and reduce flexibility to update the states of dependent activities. In the following subsections, we discuss these challenges.

\subsection{\textbf{Multiple Dependency Paths: Non-deterministic or Deterministic?}}

A requested activity may depend on a single or multiple activities in any phase of its life cycle. Multiple dependency paths (as shown in Figure \ref{fig-multiple_dependency_path} (a) and (b)) can lead to increased complexity in determining the path which the system should take first. On a different note, the order of dependency checks and updates (if required) can raise the question of whether the selection of order should be deterministic or non-deterministic. We define the deterministic and non-deterministic order of dependency check and update and later in this section, we explain which strategy is chosen for the selection of dependency path.

\begin{itemize}[leftmargin=*]
  \item \textbf{Deterministic order of dependency check and update:}  In a deterministic order for checking and updating the dependent activity states across multiple dependency paths, we can enforce a very specific selection criteria based on which order of dependency checks among a finite number of dependent activities is determined. In Figure \ref{fig-multiple_dependency_path} (a) and (b), we show two examples of activity dependency chains where act$_1$ has three dependent activities, thus it has multiple dependency paths.
  In Figure \ref{fig-multiple_dependency_path} (b), act$_6$ is a common dependency for both act$_2$ and act$_3$. For example, we can say that the current state of act$_6$ is "running" and to resolve act$_2$,  act$_6$ has to be in the "inactive" state. Moreover, act$_2$ needs act$_6$ to stay in the "inactive" state. On the other hand, to resolve act$_3$, the desired state of act$_6$ is "finished". Conceptually, according to the life cycle of an activity, it goes to an "inactive" state from "finished" state after a certain time if there are no further dependencies (post-dependent activities). We consider such a scenario for act$_6$ in \ref{fig-multiple_dependency_path} (b). If we choose one of these two dependency paths,  act$_1 \longrightarrow$ act$_3 \longrightarrow$ act$_5 \longrightarrow$  act$_6$ and act$_1 \longrightarrow$ act$_3 \longrightarrow$ act$_6 \longrightarrow$  act$_5$  starting from act$_1$ followed by act$_3$, act$_6$ will get the state "finished" and it will go to the "inactive" state since there are no other dependencies required to be checked for act$_6$. As a consequence, act$_2$ can be resolved as it can have the act$_6$ in the desired state "inactive" while checking its dependencies. This dependency check and update process is deterministic as we select the starting path comparing two different states of a common dependent activity. This selection also results in the expected outcome by resolving the chain of dependencies. However, this deterministic solution can be difficult to apply to accomplish the ultimate goal where there exists a large number of activities with multiple dependency paths including common dependent activities with different desired states.
    \item \textbf{Non-deterministic order of dependency check and update:} The non-deterministic approach for dependency check and update refers to the strategy where the sequence of activity dependency checks and updates is not fixed as well as unpredictable if an activity has multiple dependency paths. Evaluation of dependencies and update process can vary in the order each time the dependencies are checked for a specific activity. In Figure \ref{fig-multiple_dependency_path} (a) and (b), act$_1$ has three dependent activities, thus it has multiple dependency paths. In a non-deterministic selection of dependency path, the criteria to select the order of checking and updating the states of dependent activities (if required) is not predefined by the system. It can be randomly chosen and the external system does not have access to know the selection process.
\end{itemize}







In Figure \ref{fig-multiple_dependency_path} 
 (a) and (b), we show six activities in the circles named act$_1$, act$_2$, act$_3$, act$_4$, act$_5$, and act$_6$. act$_1$ is the requested activity, thus we can refer to it as the root of the dependency chain. Both the (a) and (b) in Figure \ref{fig-multiple_dependency_path} include act$_2$, act$_3$, and act$_4$ as dependent activities of act$_1$. For instance, we can think of these three activities as pre-dependent activities of act$_1$ which means we need these three activities in their respective desired states before starting act$_1$. The difference between (a) and (b) in Figure \ref{fig-multiple_dependency_path} is the parent activities of act$_6$. In Figure \ref{fig-multiple_dependency_path}(a), act$_3$ depends on act$_6$ along with act$_5$ whereas in Figure \ref{fig-multiple_dependency_path} (b) both the act$_2$ and act$_3$ depend on act$_6$ for their state change into the respective desired states. In the first figure \ref{fig-multiple_dependency_path}(a), there is no common dependency which means every dependent activity has only one parent activity in the dependency chain. On the contrary, in \ref{fig-multiple_dependency_path}(b), act$_6$
is a common dependent activity for both act$_2$ and act$_3$. For the first instance in \ref{fig-multiple_dependency_path}(a), there is no complex situation while resolving the chain of dependencies since all the dependent activities can change their current state to the desired state (if required) for their corresponding parent activities. Thus, the order of evaluating the  
dependencies and the update process does not matter in this scenario. Therefore, whether we choose deterministic or non-deterministic approach for dependency checks and update for dependency chains does not matter where there are no common dependencies between two or more parent activities.

In Figure \ref{fig-multiple_dependency_path} (b), in the dependency chain of the requested activity (act$_1$), 
 act$_2$, and act$_3$, both depend on act$_6$ in order to change their current state to the respective desired states. In this instance, there can be one of the two possible cases; requiring the same desired state of act$_6$ for both of these activities (act$_2$ and act$_3$) or requiring different desired states of act$_6$ for each activity. There does not exist any conflict if act$_6$ requires to be in the same desired state in order to change states (to their desired ones) of act$_2$ and act$_3$. However, conflict will arise when act$_2$ and act$_3$ require two different desired states for act$_6$. We refer to these different desired states as "conflicting desired states". 
In scenarios where a dependent activity has conflicting desired states, we may choose deterministic order of dependency check and update that  can provide an ultimate result where the root activity (act$_1$ in Figure \ref{fig-multiple_dependency_path} (b)) can certainly make its transition to the desired state. However, we cannot guarantee the expected outcome for the root activity of the dependency chain even if we take a deterministic solution. For example, we can compare the conflicting desired states of the common dependent activity (act$_6$) and take the most preceding state among those different desired states so that the common dependent activity can get a scope to transition to the next desired states. We need to backtrack to determine the order of the paths from the root activity to the common dependent activity with the most preceding desired state. However, we may not be able to get the desired outcome in this deterministic solution if the common dependent activity (act$_6$) needs to remain in a specific state (e.g. "inactive") to change the current state of one the parent activities (e.g. act$_2$). Moreover, act$_2$ needs to hold the specific desired state to change its state first, and then act$_1$ (root activity) can change its state. On the other hand, act$_3$ needs to change the state of act$_6$ to "running" state in order to change the state of its parent activity (root activity "act$_1$"). In this case, act$_2$ will be able to hold (which we also refer to "lock") act$_6$ in the desired state of "inactive", thus act$_3$ cannot change it to the "running" state. This is a policy conflict that cannot be solved either we choose a deterministic or non-deterministic approach and it certainly cannot provide any desired outcome for the root activity (act$_1$.). This is a policy design issue that should be handled while designing the policy and must be avoided to resolve a chain of dependency with multiple dependency paths problems.

When deciding about the deterministic approach to resolve the chain of dependencies, it becomes more complex when there are multiple levels of dependencies including activities with multiple desired states. Finding the specific order for every single activity chain is not flexible and scalable. Therefore, the system may choose the order of dependency check and update and we can leave it as a non-deterministic approach. However, choosing a non-deterministic order may sometimes lead to a race condition state. In the following section, we will address this problem and provide a solution for it. 

\subsubsection{\textbf{Race Condition Problem with Non-deterministic Order of Dependency Check and Updates with Multiple Desired States}} 
In non-deterministic execution order, we need to make sure that the state of a common dependent activity with conflicting desired states cannot be overwritten or updated when its parent activity (in the selected path from multiple dependency paths using non-deterministic order) needs the common dependent activity in a specific state. Since activity is a long continuous event, there may exist a scenario where the dependent activity fulfills the requirement and later, it can change the state according to the system context and design. Here, the race condition refers to "racing" to modify the common dependent activity's state by multiple parent activities. We need to make sure the system does not allow a parent activity to change the common dependent activity's state while another parent activity wants it to stay in another conflicting state. This race condition formulates a problem of how the system can handle the situation where a parent activity holds a dependent activity with conflicting (multiple) desired states in a specific state for a certain duration and this state cannot be overwritten by any other activity at the same time. We propose a solution using the following steps.
\setlength{\textfloatsep}{0pt}
\begin{algorithm*}[!htb]
\caption{Detecting Conflicting Desired States of Dependent Activities}
\label{alg:algorithm1}

\begin{algorithmic}

\State \textbf{RECURSIVE-CHECK-OF-DEPENDENCIES-WITH-CONFLICT-DETECTION}(\textit{da, da\_current\_st, da\_desired\_st)}:
\State \textbf{Description}: detects conflicting desired states for a chain of dependent activities.
\State \textbf{Input}: \textbf{$da$}: a dependent activity\\  \;\;\;\;\;\;\;\;\;\;\;\textbf{$da\_current\_st$}: the current state of the dependent activity $da$\\ \;\;\;\;\;\;\;\;\;\;\;\textbf{$da\_desired\_st$}: the desired state of the dependent activity, $da$.
\end{algorithmic}

\begin{algorithmic}[1]

\State \textbf{DETECT-CONFLICTING-DESIRED-STATE}(\textit{da, da\_desired\_st})
\State \textit{DoDA = $getDoDA(da,da\_current\_st,da\_desired\_st)$}
\State \textbf{if} (\textit{DoDA} $\neq \emptyset$)
\State \textbf{then}
\For{(each \textit{doda} $\in$ \textit{DoDA})}
 \State \;\;\;\;\textbf{RECURSIVE-CHECK-OF-DEPENDENCIES-WITH-CONFLICT-DETECTION}($doda$,$getCurrentSt(doda)$,\\
\hspace*{.32in}\textit{$getDesiredDoDASt(da,da\_current\_st,da\_desired\_st,doda)$})
\EndFor 
\State \textbf{end if}
\end{algorithmic}
\begin{algorithmic}
\State \textbf{\textbf{DETECT-CONFLICTING-DESIRED-STATE}}($da, da\_desired\_st$):
\State \textbf{Description}: detects conflicting desired states and stores the information for a dependent activity.
\State \textbf{Input}: \textbf{$da$}: a dependent activity\\  \;\;\;\;\;\;\;\;\;\;\;\textbf{$da\_desired\_st$}: the desired state of the dependent activity $da$
\end{algorithmic}
\begin{algorithmic}[1]
\State \textbf{if} ($assignedDesiredSt(da)$ == $\emptyset$)    
\State \textbf{then} $assignedDesiredSt(da) = da\_desired\_st$\\
\;\;\;\;\;\;\;\;$hasConflictingDesiredSt(da) =$ FALSE
\State \textbf{else if} $da\_desired\_st \neq assignedDesiredSt(da)$
\State \textbf{then} $hasConflictingDesiredSt(da) = $TRUE
\State \textbf{end if}
\end{algorithmic}
\end{algorithm*}
\setlength{\textfloatsep}{0pt}
\begin{itemize}[leftmargin=*]
    \item Initially, we check whether there exist conflicting desired states (multiple) for the dependent activities in a chain of dependencies. We store this information for future usage (referred to as Algorithm \ref{alg:algorithm1}).
    \item We introduce a recursive update process for dependent activities (in Algorithm \ref{alg:algorithm2}) where it completes the updates if the dependent activities fulfill the requirements of desired states. If there are conflicting desired states for a dependent activity, we use the locking mechanism (Algorithm \ref{alg:algorithm3}) for the dependent activity. The lock remains until the parent activity's purpose is served. 
\end{itemize}

Algorithm \ref{alg:algorithm1} is utilized to determine whether a dependent activity possesses conflicting desired states, where multiple parent activities require different desired states for the dependent activity. This algorithm consists of two functions: \textbf{RECURSIVE-CHECK-OF-DEPENDENCIES-WITH-CONFLICT-DETECTION}($da, da\_current\_state, da\_desired\_state$) and \textbf{DETECT-CONFLICTING-DESIRED-STATE}($da, da\_desired\_state$). In the \textbf{RECURSIVE-CHECK-OF-DEPENDENCIES-WITH-CONFLICT-DETECTION}($da, da\_current\_state, da\_desired\_state$) function, $da$ represents the dependent activity for which conflicting desired states are detected, with $da\_current\_state$ representing its current state and $da\_desired\_state$ denoting the desired state of the dependent activity. The function \textbf{DETECT-CONFLICTING-DESIRED-STATE}($da, da\_desired\_state$) is employed to identify conflicting desired states. It takes the dependent activity $da$ and the currently examined desired state ($da\_desired\_state$) as inputs. In this function, line 1 checks if a desired state is already assigned to $da$ using the $assignedDesiredState(da)$ function. If the function returns an empty set ($\emptyset$), we assign the currently examined desired state, $da\_desired\_state$, as the result. At this stage, as no other desired state has been checked for $da$, we can infer that no conflicting desired state exists for $da$ and assign "FALSE" as the result of $hasConflictingDesiredState(da)$. However, if there is a difference between the currently examined desired state $da\_desired\_state$ and the assigned Desired State ($assignedDesiredState(da)$), we conclude that the dependent activity $da$ possesses conflicting desired states and assign "TRUE" as the result of $hasConflictingDesiredState(da)$. After executing line 1 (calling \textbf{DETECT-CONFLICTING-DESIRED-STATE}($da, da\_desired\_state$)), we identify conflicting desired states for "dependencies of dependencies." In line 2, we obtain the set of dependent of dependent activities, $DoDA$, for the dependent activity $da$. Line 3 checks if this set is empty. If $DoDA$ is not empty, we recursively call the \textbf{RECURSIVE-CHECK-OF-DEPENDENCIES-WITH-CONFLICT-DETECTION} function for each "dependent of dependent" activity to detect conflicting desired states for these activities.

\begin{algorithm*}[!htb]
\caption{Recursive Update of States for Chain of Dependent Activities}
\label{alg:algorithm2}

\begin{algorithmic}
\State \textbf{RECURSIVE-UPDATE}(\textit{da, da\_current\_st, da\_desired\_st}):
\State \textbf{Description}: Recursively updates the states of dependent activities while exploring the dependencies of dependencies and updating them first.
\State \textbf{Input}: \textbf{$da$}: a dependent activity\\
\;\;\;\;\;\;\;\;\;\;\;\textbf{$da\_current\_st$}: the current state of the dependent activity $da$\\
\;\;\;\;\;\;\;\;\;\;\;\textbf{$da\_desired\_st$}: the desired state of the dependent activity, $da$.
\State \textbf{Output}: Returns a desired state for the dependent activity, $da$.
\end{algorithmic}

\begin{algorithmic}[1]
\State \textit{DoDA = $getDoDA(da,da\_current\_st,da\_desired\_st$)} 
\State \textbf{if} (\textit{DoDA} == $\emptyset $)
\State \textbf{then return} \textit{da\_desired\_st};
    \State \textbf{else}
\For{(each $doda \in DoDA$)}
\State \textbf{if} \textit{($getCurrentSt(doda)$ $\neq$ $getDesiredDoDASt(da,da\_current\_st,da\_desired\_st,doda$)}\\ 
\;\;\;\;\;\;\;\;\; $\land$ \textit{hasConflictingDesiredSt(doda)} == FALSE) 
\State \textbf{then}
\State \;\;\;\;\; \textit{$getCurrentSt(doda)$ = \textbf{RECURSIVE\_UPDATE}(doda, 
  $getCurrentSt(doda)$, }
\State \;\;\;\;\;\textit{ $getDesiredDoDASt(da, da\_current\_st, da\_desired\_st, doda)$)}
 
\State \textbf{else if} \textit{($getCurrentSt(doda)$ $\neq$ $getDesiredDoDASt(da,da\_current\_st,da\_desired\_st,doda$)} 
 \State \;\;\;\;\;\;\;\;\;\;$\land$ \textit{$hasConflictingDesiredSt(doda)$} == TRUE)
\State \textbf{then} \State \;\;\;\;\;\; \textbf{ACQUIRE-LOCK}($da, da\_current\_st, da\_desired\_st,  doda, getBinarySemaphoreValue(doda)$,\State \;\;\;\;\;\; $getDesiredDoDASt(da, da\_current\_st, da\_desired\_st, doda))$
 \State \;\;\;\;\;\;\;\textit{$getCurrentSt(doda)$ = \textbf{RECURSIVE\_UPDATE}(doda, 
 $getCurrentSt(doda)$,}
\State \textit{\;\;\;\;\;\;\;$getDesiredDoDASt(da, da\_current\_st, da\_desired\_st, doda)$)}\;\;\;\;\;\;\Comment{\texttt{\textbf{RELEASE-LOCK}(doda) will be called when the purpose of locking $doda$ is done for $da$}}
\State \;\;\textbf{end if}
\EndFor
\State \textbf{return $da\_desired\_st$}
\State \textbf{end if}
\end{algorithmic}
\end{algorithm*}
\setlength{\textfloatsep}{0pt}
\begin{algorithm*}[!htb]
\caption{Locking Mechanisms for Activities with Conflicting Desired States}
\label{alg:algorithm3}

\begin{algorithmic}

\State \textbf{\textbf{ACQUIRE-LOCK}}($da$, $da\_current\_st, da\_desired\_st$ , $doda$, $getBinarySemaphoreValue(doda)$\\$getDesiredDoDASt(da, da\_current\_st, da\_desired\_st, doda)$):
\State \textbf{Description}: Lock a dependent of dependent activity if it is unlocked and wait for the release of lock if it is locked.
\State \textbf{Input}: \textbf{$da$}: a dependent activity,\\ \;\;\;\;\;\;\;\;\;\;\;\textbf{$da\_current\_st$}: the current state of the dependent activity $da$,\\ \;\;\;\;\;\;\;\;\;\;\;\textbf{$da\_desired\_st$}: the desired state of the dependent activity, $da$.\\
\;\;\;\;\;\;\;\;\;\;\;\textbf{$doda$}: a dependent of dependent activity.\\
\;\;\;\;\;\;\;\;\;\;\;\textbf{$getBinarySemaphoreValue(doda)$}: the binary semaphore value of $doda$ which can be 0 or 1 in turn.\\
\;\;\;\;\;\;\;\;\;\;\;$getDesiredDoDASt(da, da\_current\_st, da\_desired\_st, doda)$: the desired state of $doda$ corresponding to the\\
\;\;\;\;\;\;\;\;\;\;\;parent activity $da$'s state transition from $da\_current\_st$ to $da\_desired\_st$.

\end{algorithmic}

\begin{algorithmic}[1]
\State \textbf{if}($getBinarySemaphoreValue(doda)$==1)
\State \textbf{then} $getBinarySemaphoreValue(doda)$ = 0;

\State \textbf{else if} ($getBinarySemaphoreValue(doda)$==0)
\State \textbf{then} 
\;\;\;\;\While{($getBinarySemaphoreValue(doda)$==0)} 
\;\;\;\;\;\;\;\;\State \textit{waitFor(doda)}
\EndWhile
\State \textbf{ACQUIRE-LOCK}($da, da\_current\_st, da\_desired\_st, 
 doda, 
 getBinarySemaphoreValue(doda)$,\State 
 $getDesiredDoDASt(da, da\_current\_st, da\_desired\_st, doda))$
\State \textbf{end if}
\end{algorithmic}

\begin{algorithmic}

\State \textbf{\textbf{RELEASE-LOCK($doda$):}}
\State \textbf{Description}: releases the lock for a dependent of dependent activity.
\State \textbf{Input}: \textbf{$doda$}: a dependent activity
\end{algorithmic}

\begin{algorithmic}[1]
\State $getBinarySemaphoreValue(doda)$ = 1;
\end{algorithmic}

\end{algorithm*}

Algorithm \ref{alg:algorithm2} is implemented to handle the recursive process of checking and updating the states of dependent activities within a dependency chain. We choose a recursive structure for this algorithm to ensure that we address the "dependencies of dependencies" before updating the state of a dependent activity. The function \textbf{RECURSIVE-UPDATE}($da, da\_current\_st, da\_desired\_st$) is defined, where $da$ represents a dependent activity, $da\_current\_st$ denotes its current state, and $da\_desired\_st$ denotes the desired state for $da$.
In line 1, we obtain the set of dependent activities ($DoDA$) that are required for $da$ to transition from $da\_current\_st$ to $da\_desired\_st$. If $DoDA$ is empty, it implies that there are no dependencies that need to be checked for this specific state transition of $da$. Line 2 verifies whether $DoDA$ is empty or not. 
If the condition is true, we return $da\_desired\_st$ from the function (line 3). Otherwise (line 5), we proceed to explore each activity ($doda$) in $DoDA$.
Within lines 6-7, we check whether the current state and desired state of each $doda$ in $DoDA$ are not the same and whether $doda$ has any conflicting state. This information has already been stored using Algorithm \ref{alg:algorithm1} for all activities in the dependency chain. If the condition in lines 6-7 is met, we update the state of $doda$ by calling the \textbf{RECURSIVE-UPDATE} function, providing $doda$, its current state, and desired state as parameters (lines 8-10). This recursive call is necessary to check if $doda$ has any further dependent activities and to compare their current and desired states before returning the desired state.
We include an additional check to verify if the current and desired states of $doda$ are not equal and if $doda$ has any conflicting states (lines 11-12). 
If these conditions are satisfied,
we call \textbf{ACQUIRE-LOCK} function defined in Algorithm \ref{alg:algorithm3} (line 14-15). Algorithm \ref{alg:algorithm3} will check if the activity, $doda$ is locked or unlocked. Then we update the current state of this activity by calling the function  \textbf{RECURSIVE-UPDATE} (line 16-17).

Algorithm \ref{alg:algorithm3} is inspired by the Binary Semaphore \cite{cho2010low} or Mutex Lock mechanisms in operating systems. The binary Semaphore mechanism is used to synchronize between two values, 0 and 1, and allows only a single unit to the critical section (to get access to shared resources). We use a similar locking mechanism using a function named "$getBinarySemaphoreValue(doda)$" where $doda$ is a dependent of dependent activity and the value returned from this function is 0 or 1. When the value returned from "$getBinarySemaphoreValue(doda)$" is 1, this indicates that $doda$ is currently not locked by a parent activity. When the value returned from "$getBinarySemaphoreValue(doda)$" is 0, this indicates $doda$ is currently locked by a parent activity. Therefore, it cannot change its current state to fulfill the requirement of any other parent activity. In the function \textbf{\textbf{ACQUIRE-LOCK}}($da$, $da\_current\_st, da\_desired\_st$, $doda$, $getBinarySemaphoreValue(doda)$, $getDesiredDoDASt(da, da\_current\_st, da\_desired\_st, doda)$), $da$ is the dependent activity which is currently trying to change the current state of $doda$ to the desired state in order to transition from $da\_current\_st$ to $da\_desired\_st$. 
In this algorithm, we check if $doda$ is currently locked ($getBinarySemaphoreValue(doda)$ == 0) or unlocked ($getBinarySemaphoreValue(doda)$== 1). If it is unlocked, we change the value of $getBinarySemaphoreValue(doda)$ to 0 which indicates it is locked by $da$ (line 1-2). If $doda$ is locked by some other activity (line 3), $da$ must wait for $doda$ to be unlocked until we get the value 1 from $getBinarySemaphoreValue(doda)$ (line 5-7). $waitFor(doda)$ indicates, the parent activity $da$ will wait for $doda$ to be unlocked. Once the previous parent activity releases the lock using \textbf{RELEASE-LOCK}($doda$) and changes the value of $getBinarySemaphoreValue(doda)$ to 1, the currently waiting dependent activity $da$ again calls the \textbf{ACQUIRE-LOCK} function and updates the value of $getBinarySemaphoreValue(doda)$ (locks $doda$) (line 8-9).

\subsection{\textbf{Circular Dependencies and Deadlock}}
\begin{figure*}[!htb]
    \centering
\includegraphics[width=0.35\textwidth]{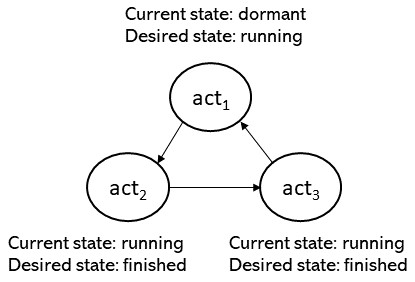}
    \centering
    \caption{Circular Dependencies of Activities.}
    \label{fig-deadlock}
\end{figure*}

In our proposed $\mathrm{ACAC_{D}}$ model, there can be circular dependencies that create a deadlock situation. In a circular set of dependencies, the chain of dependencies is created in a circular fashion (shown in Figure \ref{fig-deadlock}). In this figure, the dependency path is 
act$_1 \longrightarrow$ act$_2 \longrightarrow$ act$_3 \longrightarrow$ act$_1$. In this circular set of activities, act$_1$ depends on 
act$_2$ to find act$_2$ in the desired state "finished". act$_2$ depends on act$_3$ and requires act$_3$ to be "finished". act$_3$ requires act$_1$ to be "running" before it goes to "finished" state. This circular wait for each activity is in a deadlock and no activity ultimately gets their desired state.
There are certain ways to handle this type of deadlock situations. We discuss the deadlock handling techniques in the next subsection.


\subsubsection{Deadlock Detection and solutions}
The system may fall into a deadlock if the system administrator fails to prevent it from assigning the dependencies that create a circle. The deadlock due to a circular set of dependent activities can be detected before the update process starts. We can detect this deadlock with a typical Depth First Search (DFS) algorithm. This kind of deadlock situation needs to be carefully analyzed by the system designer. Upon identifying a circular dependency in the system, the administrator plays a crucial role in breaking the cycle within the dependency chain. It becomes imperative for the administrator to thoroughly examine the activities involved in the cycle and pinpoint a low-priority activity that can be strategically removed. If the administrator successfully accomplishes this task, the deadlock can be effectively eliminated, ensuring the system's smooth operation without violating the safety rules. Therefore, careful analysis and decision-making by the administrator are instrumental in resolving such deadlock scenarios, ultimately optimizing the system's performance and preventing potential disruptions. Maximum timeout mechanisms can also be applied for a requested activity where the request is denied after a certain period of time. Deadlock detection and recovery are challenging for a chain of dependent activities since this is a design choice and policy engineering problem. Deadlock prevention is a more suitable deadlock handling method for a chain of dependent activities in such smart systems referred to by our proposed ACAC model.

\subsection{Combination of $\mathrm{ACAC_{D}}$ Sub Models while Resolving Chain of Dependencies}

As described in Section \ref{Formal_ACAC}, different sub-models, denoted as $\mathrm{ACAC_{D}}$ sub-models, support the mutability of activities at various stages of their life cycle. Throughout the paper, we discuss several states ($inactive, dormant, aborted, running, hold, revoked, finished$) that a requested activity can pass through from its initiation to completion. To fulfill the request for an activity, the states of dependent activities are examined and updated, if necessary, to enable the transition of the requested activity from one state to another. For instance, when transitioning from the \textit{inactive} to \textit{running} state, the $\mathrm{ACAC_{preD_1}}$ model is employed to verify and modify the states of the dependent activities required to initiate the requested activity. During the \textit{running} state of the requested activity, the $\mathrm{ACAC_{onD_2}}$ model is utilized for conducting ongoing checks and updates. Consequently, it can be inferred that a combination of different $\mathrm{ACAC_{D}}$ sub-models supports the successful execution of a requested activity from start to finish.

A "chain of dependent activities," also known as "dependencies of dependencies", requires checks for current and desired states of the dependent activities and updates to the current states to accommodate state transitions of parent activities. As illustrated in Figure \ref{fig-chain_of_dependencies_example} in Section \ref{Formal_ACAC}, consider the example where the activity "Water Spraying" requires "Nitrogen Spraying" to be in an \textit{inactive} state while it is currently in a \textit{running} state. Similarly, the state transition of "Mixing Sawdust to soil" from \textit{running} to \textit{inactive} is required. Suppose "Nitrogen Spraying" first transitions to a \textit{finished} state before reaching the \textit{inactive} state. Since it has no post-dependent activities, it will go to \textit{inactive} state. In this scenario, "Mixing Sawdust to soil" is checked and updated when "Nitrogen Spraying" is ongoing, and a decision is made to finish the ongoing activity. Based on the definitions of our $\mathrm{ACAC_{D}}$ models, $\mathrm{ACAC_{onD_2}}$ model can be utilized to check and update the "Mixing Sawdust to soil" activity, while $\mathrm{ACAC_{onD_3}}$ model ensures there are no post-dependent activities. Hence, combination of $\mathrm{ACAC_{D}}$ sub-models proves to be an effective approach for resolving the chain of dependencies.\\
\vspace{-4mm}

\section{Prototype Implementation}
\label{sec:implementation}
In this section, we present a prototype implementation of a combination of $\mathrm{ACAC_{D}}$ sub-models in a smart farming use case (as shown in Figure \ref{dependent_activities}). The code is written in Python 3 using PyCharm on Hp Envy x360 convertible with Intel core i7 processor and 12 GB of RAM. The implementation shows the need for different $\mathrm{ACAC_{D}}$ sub-models to incorporate the dependencies (D) in the activity request decision. In a fully deployed ACAC model, all four decision parameters (Authorizations (A), Obligations (B), Conditions (C), and Dependencies (D)) will be considered. However, since our paper focuses on the $\mathrm{ACAC_{D}}$ models for activity dependencies, we evaluate these sub-models, assuming other parameters are satisfied. We have simulated the devices and activities in the system; however, this does not undermine the plausibility, use, and advantage of our proposed $\mathrm{ACAC_{D}}$ model, as further elaborated in the following discussion.



\begin{table}[!t]
\setlength{\tabcolsep}{2pt}
\renewcommand{\arraystretch}{0.5}
\centering

\caption{Description of activity requests and dependencies}
\vspace{-2mm}
\scalebox{0.8}{%
\begin{tabular}{*{15}{|p{2.7cm}|p{4.5cm}|p{4.2cm}|p{3.6cm}}}
\hline

\textbf{Requests} & \textbf{Pre-dependent activities}&\textbf{Ongoing-dependent activities}& \textbf{Post-dependent activities}\\
\hline
\textit{request(fieldWorker,
sprayingWeedKiller)}&
\vspace{-1.5mm}
 \begin{itemize}[noitemsep,nolistsep,leftmargin=*]
  \item \textit{mixingAMS} : \textit{finished}
  \item \textit{thermalImaging} : \textit{running}
\end{itemize}
\vspace{-2mm}&
\vspace{-1.5mm}
 \begin{itemize}[noitemsep,nolistsep,leftmargin=*]
  \item \textit{waterSpray} : \textit{inactive}
  \item \textit{thermalImaging} : \textit{running}
\vspace{-1.7mm}
\end{itemize}
\vspace{-1.5mm}&
\vspace{-1.5mm} \begin{itemize}[noitemsep,nolistsep,leftmargin=*]
  \item \textit{waterSpray} : \textit{inactive}
  \item \textit{pullingWeedsUp} : \textit{running}
  \vspace{-1.5mm}
\end{itemize}
\\
\rowcolor{lightgray!40!}
\hline
\textit{request(farmer, sowingSeeds)}&
\vspace{-1.5mm} \begin{itemize}[noitemsep,nolistsep,leftmargin=*]
  \item \textit{fieldPloughing} : \textit{inactive}
\end{itemize}&
\vspace{-0.7mm} \begin{itemize}[noitemsep,nolistsep,leftmargin=*]
\vspace{-1.5mm}
  \item \textit{pesticideSpray} : \textit{running}
  \item \textit{thermalImaging} : \textit{running}
  \item \textit{airCooling} : \textit{running}
\vspace{-1.7mm}
\end{itemize}
\vspace{-1.5mm}&
N/A
 \\
 \hline
\textit{request(farmManager, fieldPloughing)}&
 \begin{itemize}[noitemsep,nolistsep,leftmargin=*]
\vspace{-1.5mm}
  \item \textit{stakingBoundaries} : \textit{finished}
  \item\textit{mixingWaterAbsorbingMaterial} : \textit{running}
  \vspace{-1.9mm}
\end{itemize}&
\vspace{-1.7mm} \begin{itemize}[noitemsep,nolistsep,leftmargin=*]
  \item \textit{waterSpray} : \textit{inactive}
  \item \textit{thermalImaging} : \textit{running}
  \vspace{-1.5mm}
\end{itemize}&
\vspace{-1.5mm} \begin{itemize}[noitemsep,nolistsep,leftmargin=*]
  \item \textit{sprayingWeedKiller} : \textit{running}
  \item \textit{sowingSeeds} : \textit{running}
  \item \textit{pesticideSpray} : \textit{running}
  \vspace{-1.7mm}
\end{itemize}
 \\
  \rowcolor{lightgray!40!}
 \hline
\textit{request(fieldOwner, coolingGreenhouse)}&
\vspace{-0.7mm} \begin{itemize}[noitemsep,nolistsep,leftmargin=*]
\vspace{-1.7mm}
  \item \textit{thermalImaging} : \textit{running}
\end{itemize}&
\vspace{-2mm} \begin{itemize}[noitemsep,nolistsep,leftmargin=*]
  \item \textit{humidifying} : \textit{running}
\end{itemize}&N/A
 \\
\hline

\end{tabular}

}
\label{request}
\end{table}

\begin{table}[!t]
\setlength{\tabcolsep}{2pt}
\renewcommand{\arraystretch}{0.5}
\centering

\caption{Chain of dependencies for the dependent activities in the first request from Table \ref{request}}
\vspace{-2mm}
\scalebox{0.8}{%
\begin{tabular}{*{15}{|p{2.7cm}|p{4.5cm}|p{4.2cm}|p{3.6cm}}}
\hline

\textbf{Dependent activity} & \textbf{Current state}&\textbf{Desired state}& \textbf{Dependent of dependent activity: desired State}\\
\rowcolor{lightgray!40!}
\hline
\textit{mixingAMS}&
\vspace{-1.5mm}
 \textit{running}
\vspace{-2mm}&
\vspace{-1.5mm}
\textit{finished}
\vspace{-1.5mm}&
\vspace{-1.5mm} \begin{itemize}[noitemsep,nolistsep,leftmargin=*]
  \item \textit{mixingVinegar} : \textit{running}
  \vspace{-1.5mm}
\end{itemize}
\\
\hline
\textit{pullingWeedsUp}&
\vspace{-1.5mm}
 \textit{inactive}
\vspace{-2mm}&
\vspace{-1.5mm}
\textit{running}
\vspace{-1.5mm}&
\vspace{-1.5mm} \begin{itemize}[noitemsep,nolistsep,leftmargin=*]
  \item \textit{pesticideSpray} : \textit{running}
  \vspace{-1.5mm}
\end{itemize}
 \\
 \rowcolor{lightgray!40!}
 \hline
 \textit{mixingVinegar}&
\vspace{-1.5mm}
 \textit{inactive}
\vspace{-2mm}&
\vspace{-1.5mm}
\textit{running}
\vspace{-1.5mm}&
\vspace{-1.5mm} \begin{itemize}[noitemsep,nolistsep,leftmargin=*]
  \item \textit{mixingWater} : \textit{running}
  \vspace{-1.5mm}
\end{itemize}
 \\
 
\hline

\end{tabular}

}
\label{chain_of_dependencies_for_request}
\end{table}

\subsection{Description of the Use Case}
A smart farming ecosystem consists of connected smart devices that perform multiple activities concurrently. There are inter-dependencies among activities that may constrain the execution of other activities. This requires checking and updating the states of dependent activities to make any activity request decision. 
In Table \ref{request}, we include four activity requests in the first column. Each request has two parameters; the first and second parameter indicates the requesting source and the requested activity, respectively.
The second, third, and fourth columns include \textit{pre-}, \textit{ongoing}, and \textit{post-} dependent activities, respectively. We also mention the desired states (such as \textit{running} or \textit{inactive}) of dependent activities after the colon `:'. Since the current states of the activities depend on the real-time system context, these are not specified. Further, we implement an activity request with its chain of dependencies. In Table \ref{chain_of_dependencies_for_request}, we include the dependencies of dependencies corresponding to the first request shown in Table \ref{request}. The first column indicates the name of the dependent activities, the second and third columns indicate its current and desired states respectively. The fourth column includes the dependent of dependent activities corresponding to the transition from the current state to the desired state of the parent dependent activities mentioned in the first column. We also mention the corresponding desired states of the dependent of dependent activities followed by a colon `:'.

\subsection{Use Case Implementation}
To implement the use case and satisfy the activity requests in Table \ref{request}, we configure five JSON files as follows, \texttt{request.json} (have the activity requests with a source and requested activity), \texttt{activity.json} (includes the current states of all the activities), \texttt{object.json} (holds the objects the activities can be performed on), \texttt{operation.json} (contains the operation to perform an activity on a specific object for all the activities), and \texttt{activityDependencies.json} (provides the sets of pre-, ongoing- and post-dependent activities with their desires states and against particular object for each requested activity). \texttt{activity.json} file is dynamically updated according to the changes made in the current states of dependent activities. Further, we configure another JSON file named \texttt{dependenciesOfdependencies} to implement this use case with chain of dependencies where pre-, ongoing and post-dependent activities (for a particular activity request) also have dependent activities to make their transition from the current state to a desired states while satisfying the requested activity's requirements.

As mentioned in Table \ref{request}, for the
\textit{request(fieldWorker, sprayingWeedKiller)}, we have all three of pre-, ongoing and post-dependent activities with desired states. 
The current states we get from our \texttt{activity.json} file is compared to the desired states. For this pre-dependency check, our implementation procedure supports $\mathrm{ACAC_{preD_0}}$ sub-model. The activity \textit{mixingAMS} (mixingAMS is the short form of mixingAmmoniumSulfate) is initially in \textit{running} state which needs to update its state to the desired state \textit{finished}. This update occurs in the enforcement point as supported by the $\mathrm{ACAC_{preD_1}}$ sub-model. In a similar way, the ongoing dependent activities are checked, and the current state of \textit{waterSpray} is updated from \textit{running} to \textit{inactive}. 
In this ongoing check, the sub-models $\mathrm{ACAC_{onD_0}}$ (for checking the states of ongoing dependent activities) and $\mathrm{ACAC_{onD_2}}$ (for updating the current states of the ongoing-dependent activities) are applicable. In post-check, 
a post-dependent activity, \textit{pullingWeedsUp} needs to change its state (from \textit{inactive} to \textit{running}) where the sub-model $\mathrm{ACAC_{onD_3}}$ fits the best. In summary, this use case implementation shows the combination of $\mathrm{ACAC_{preD_0}}$, $\mathrm{ACAC_{preD_1}}$ $\mathrm{ACAC_{onD_0}}$, $\mathrm{ACAC_{onD_2}}$, and $\mathrm{ACAC_{onD_3}}$ for satisfying the \textit{request(fieldWorker, sprayingWeedKiller)}. 
Similarly, for the other requests, the same procedure repeats for pre-, ongoing, post-check and thus, reflecting the applicability of our proposed $\mathrm{ACAC_{D}}$ sub-models.

\begin{table}[!htb]
\setlength{\tabcolsep}{2pt}
\renewcommand{\arraystretch}{0.85}
\centering
\caption{Execution time for pre-, ongoing, and post-check}
\vspace{-2mm}
\label{tab5:sum}
\scalebox{0.9}{%
\begin{tabular}{|c|c|c|c|c|c|c|c|c|c|}
\hline
\multirow{3}{*}{\textbf{Number of Requests}} & \multicolumn{3}{c|}{\textbf{Pre-Check}} & \multicolumn{3}{c|}{\textbf{Ongoing-Check}} & \multicolumn{3}{c|}{\textbf{Post-Check}} \\
\cline{2-10}
 & \textbf{NDC} & \textbf{NDU} & \textbf{Time} & \textbf{NDC} & \textbf{NDU} & \textbf{Time} & \textbf{NDC} & \textbf{NDU} & \textbf{Time} \\
\hline
10 & 20 & 10 & 38.84 & 30 & 10 & 42.83 & 20 & 20 & 15.21 \\
\hline
20 & 30 & 10 & 53.33 & 60 & 30 & 57.64 & 20 & 20 & 22.09 \\
\hline
30 & 50 & 0 & 101.33 & 80 & 0 & 87.24 & 50 & 10 & 25.44 \\
\hline
40 & 60 & 10 & 110.16 & 90 & 10 & 124.3 & 50 & 30 & 60.76 \\
\hline
\end{tabular}

}
\label{tab:perform1}
\vspace{-4mm}
\end{table}
\begin{table}[!htb]
\setlength{\tabcolsep}{2pt}
\renewcommand{\arraystretch}{0.85}
\centering
\caption{Execution time for pre-, ongoing, and post-check with resolving chain of Dependencies}
\vspace{-2mm}
\label{tab5:sum}
\scalebox{0.9}{%
\begin{tabular}{|c|c|c|c|c|c|c|c|c|c|}
\hline
\multirow{3}{*}{\textbf{Number of Requests}} & \multicolumn{3}{c|}{\textbf{Pre-Check}} & \multicolumn{3}{c|}{\textbf{Ongoing-Check}} & \multicolumn{3}{c|}{\textbf{Post-Check}} \\
\cline{2-10}
 & \textbf{NDC} & \textbf{NDU} & \textbf{Time} & \textbf{NDC} & \textbf{NDU} & \textbf{Time} & \textbf{NDC} & \textbf{NDU} & \textbf{Time} \\
\hline
10 & 40 & 30 & 45.89 & 30 & 30 & 32.35 & 30 & 10 & 40.14 \\
\hline
20 & 80 & 60 & 79.06 & 60 & 60 & 59.93 & 80 & 60 & 71.56 \\
\hline
30 & 120 & 90 & 116.16 & 90 & 90 & 84.56 & 120 & 90 & 99.97 \\
\hline
40 & 160 & 120 & 194.32 & 120 & 120 & 94.81 & 160 & 120 & 138.1 \\
\hline
\end{tabular}

}
\label{tab:perform2_chain}
\end{table}

To implement this use case with a chain of dependencies, we consider the first request from Table \ref{request} which is \textit{request (fieldWorker, sprayingWeedKiller)}. As mentioned in Table \ref{chain_of_dependencies_for_request}, we have the dependent of dependent activities corresponding to the transition (from current state to desired state) of parent dependent activities. To implement this request with the chain of dependencies, we configure a JSON file \texttt{dependenciesOfdependencies.json}. In pre-, ongoing, and post-check of \textit{request(fieldWorker, sprayingWeedKiller)}, the dependencies of dependencies are checked and the updates are resolved recursively where all the required updates are performed for the dependencies before its parent activity transitions to the desired state. For instance, before updating the state of \textit{pullingWeedsUp} from \textit{inactive} to \textit{running} in the post-check, we check the dependencies of dependencies and update their states accordingly if required, (\textit{pesticideSpray} updates its state from \textit{inactive} to \textit{running} for the particular required transition of  \textit{pullingWeedsUp}). The dependent activities which are not mentioned in Table \ref{tab:perform2_chain} do not have other dependent activities (DoD). In general, the dependencies that are checked when a parent activity's current state is \textit{inactive} and needs to transition to a \textit{running} state, are called pre-dependent activities. Similarly, ongoing dependencies are checked while the parent activity's current state is \textit{running} and needs the transition to any succeeding state (such as \textit{finished}, or \textit{hold}). Ongoing dependencies are also checked regularly to see whether the execution could continue or be revoked. The post-dependent activities are checked when parent activity's required state transitions are \textit{finished} or \textit{revoked} to \textit{inactive}, or \textit{hold} to \textit{running}, \textit{finished} or, \textit{revoked}.

\begin{figure}[!t]
\centering
 \includegraphics[width=0.9\columnwidth]{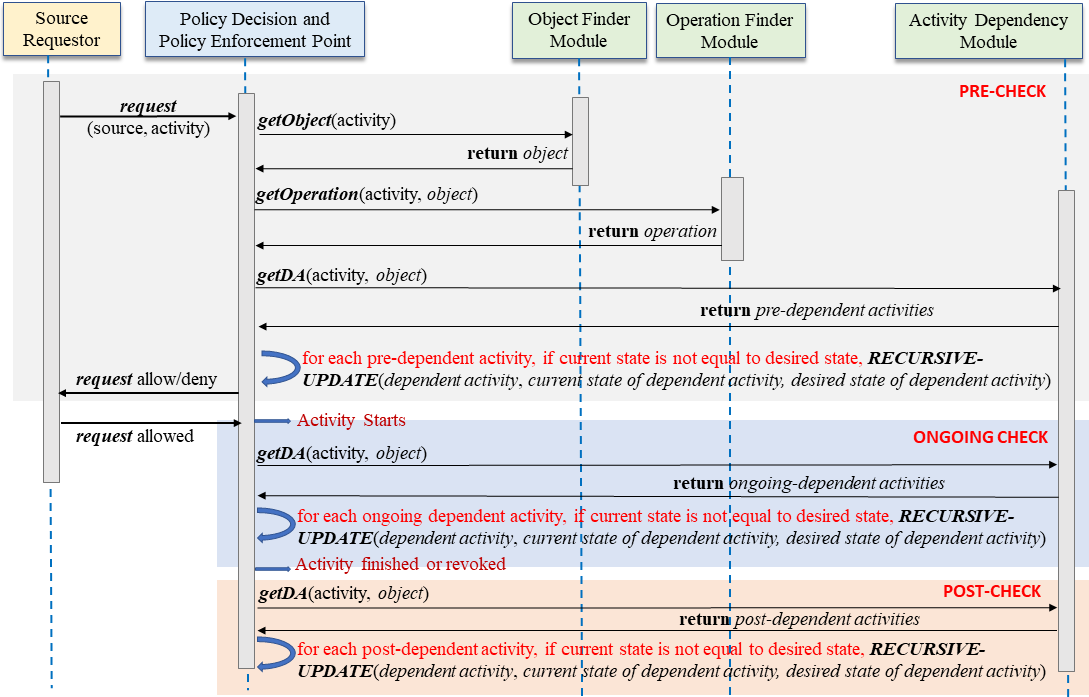}
\caption{Sequence diagram for $\mathrm{ACAC_{D}}$ Implementation}
\label{sequence_chain}
\end{figure}
The sequence of the implementation process is shown in Figure \ref{sequence_chain}. We have three phases (shown in different colors) of checking and updating the dependent activity states while satisfying the requests, referred as pre-check, ongoing check, and post-check. When a source requests an activity, it is checked at the policy decision and enforcement point, the suitable object and operation are selected (mentioned as \textit{getObject(activity) and \textit{getOperation(activity, object)}}) from the \texttt{object} and \texttt{operation} finder modules, respectively, which check the \texttt{object.json} and the \texttt{operation.json} files. In pre-check phase, the activity dependency module provides the pre-dependent activities using the \texttt{activityDependencies.json}. In the policy decision and enforcement point, for each pre-dependent activity, current and desired states are checked and updated (if required and depending on mutability). 

In our implementation without dependencies of dependencies (Table \ref{request}), the dependent activities directly update their current states without checking further dependencies. On the other side, the implementation with dependencies of dependencies (first request from Table \ref{request} and chain of dependencies of this requested activity in Table \ref{chain_of_dependencies_for_request}), in RECURSIVE-UPDATE function call (mentioned as \textit{RECURSIVE-UPDATE(dependent activity, current state of dependent activity, desired state of dependent activity)}), further dependency check (using \texttt{dependenciesOfdependencies.json} file) and the recursive update take place. 
The request is allowed or denied based on the fulfillment of the dependencies. The activity starts to run at this point. In ongoing phase, the ongoing dependent activity states are checked and updated. We assume the requested activity is finished after resolving the ongoing dependencies. Similarly, post-dependent activity states are checked and updated in the post-check after the activity is revoked or finished. The requested activity changes its current state (from \textit{finished} or \textit{revoked} to \textit{inactive}) at this point.
\subsection{Performance Evaluation}
We evaluated the implementation of our proposed $\mathrm{ACAC_{D}}$ model in different processing stages (pre-, ongoing , and post-check). 
We evaluate our prototype for the four activity requests stated in our use case by sending each activity request ten times simultaneously (assuming ten different sources request for ten different activities) and adding new requests in the same proportion.  

\begin{figure}[!htb]
\centering
 \includegraphics[width=.5\columnwidth]{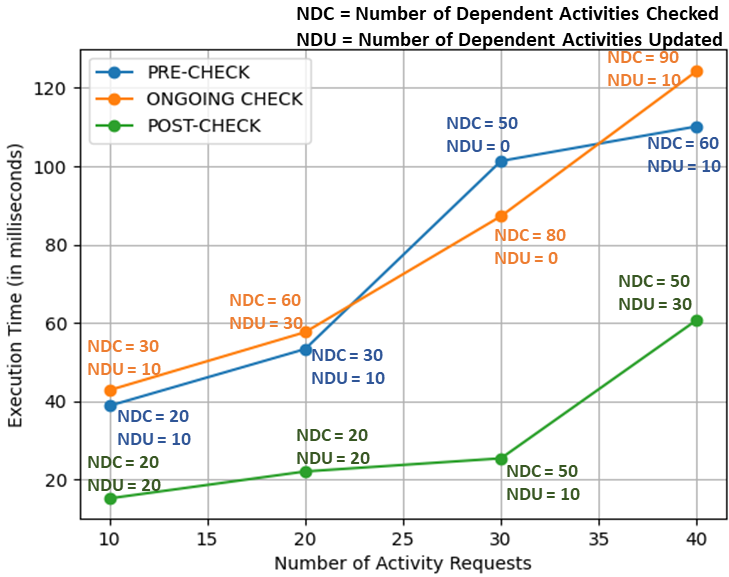}
\caption{Performance Evaluation of Implementation.}
\label{Performance_evaluation}
\end{figure} 
\begin{figure}[!htb]
\centering
 \includegraphics[width=0.55\columnwidth]{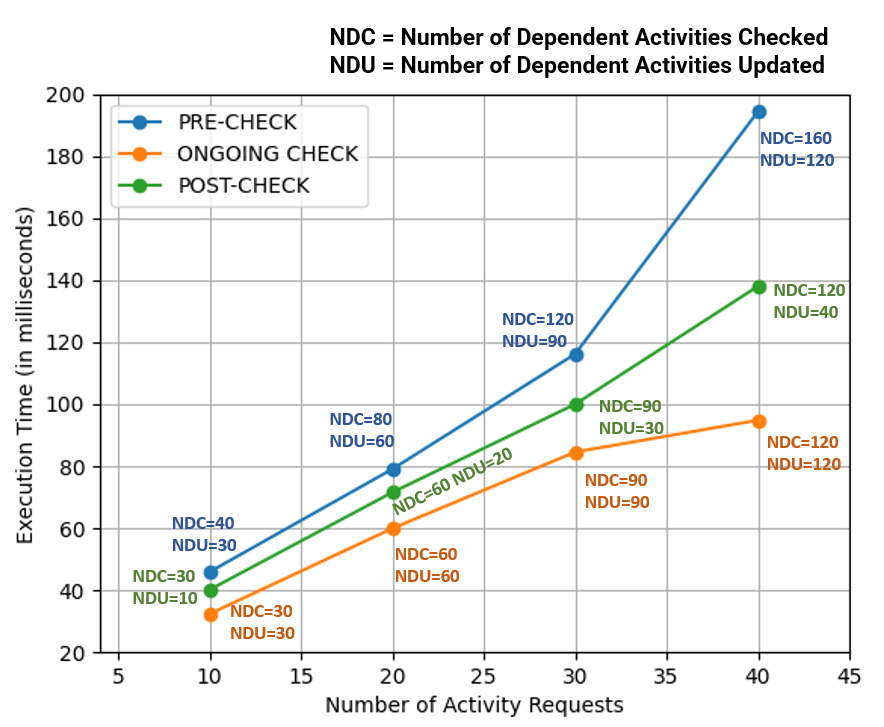}
\caption{Performance Evaluation of Implementation with Chain of Dependencies.}
\label{fig:Performance_evaluation2}
\end{figure}
Table \ref{tab:perform1} shows the execution time (in milliseconds) against the total number of requests for pre-check, ongoing check and post-check respectively. The first column indicates the number of requests. The first and second sub-columns in each of the second, third and fourth columns indicate the number of dependent activities checked (NDC) and the number of dependent activities updated (NDU), respectively in pre-check, ongoing-check and post-check.  
It must be noted that, in pre-check, the current state of a requested activity is updated from \textit{inactive} to \textit{running} if it is allowed after checking and updating the current states of the pre-dependent activities. In this case, we start the timer when the request is made and calculate the execution time until it updates the current state if the activity is allowed. In ongoing-check, after checking and updating the ongoing dependent activities, we assume the requested activity is finished and, thus, update its current state from \textit{running} to \textit{finished}. The execution time is then evaluated for the duration of the dependency checking and updating the ongoing dependent activity states (if required) and changing the current state of the ongoing requested activity from \textit{running} to \textit{finished}. The execution time for post-check indicates the duration of checking and updating (if required) the post-dependent activities. In our implementation (without chain of dependencies), the execution time of pre- and ongoing checks is more than the execution time of post-check since they perform more dependent activities' states update.
It should be noted that the number of updates on dependent activities may reduce as more activities are requested since it is possible that the earlier activity requests have already updated the states, and no more state change is needed for future requests.


Figure \ref{Performance_evaluation} compares the execution time against the number of requests considered for pre-, ongoing, and post-check (indicated by blue, red, and green lines, respectively). 
The figure shows that the execution time increases with the increase in the number of dependent activities checked and updated. We observe that the maximum calculated time is for the forty simultaneous activity requests in the ongoing check case. Since in our use case, this scenario has the maximum number of dependent activities checked along with updates to the current states of requested activities (assuming activities finished their execution). Clearly, the number of dependencies for a particular requested activity and the number of state updates impact the processing time of an activity request.

In implementing the request processing of the chain of dependencies for the first request in Table \ref{request}, we evaluate the performance by sending the same request 10, 20, 30, and 40 times. Each time, the dependencies of dependencies are checked, and their states are updated if required. This process is done in a recursive manner to ensure that dependencies are resolved before the parent activity's state changes. Table \ref{tab:perform2_chain} shows the execution time (in milliseconds) against the total number of requests (in a similar way as done in Table \ref{tab:perform1}). Figure \ref{fig:Performance_evaluation2} compares the execution time against the number of requests, similarly shown in Figure \ref{Performance_evaluation}. Here, we observe that execution time increases with the increase in total number of dependency checks and dependency updates. Since NDC and NDU are the highest in number in pre-check, the execution time is also high in pre-check.

We understand that the processing time will increase with hundreds of devices and activities running simultaneously in a real-world environment. However, in this implementation, we reflect on the plausibility and applicability of considering dependencies as a critical component to support activity control in smart systems.
\vspace{-2mm}


\section{Related Work}
\label{related_work}
With the advancement of technologies and growth in IoT devices, the possibility of violation of security mechanisms increases. Various research works, including \cite{yao2021security, babun2021survey} investigate security and privacy issues existing in smart and connected systems. 
Yao et al. describe security and privacy challenges in different working stages of physical objects in IoT \cite{yao2021security}. Access control solutions have also been proposed for the smart and automated systems, including fine-grained attribute-based access control (ABAC) \cite{ ameer2022attribute, gupta2019dynamic, chen2021blockchain, bhatt2020abac, zhang2020attribute}.


An attribute-based access control solution for industrial IoT proposed by Bhatt et al. \cite{bhatt2021attribute} implement their model in Amazon Web Services IoT. Ameer et al. proposed ABAC for secured smart home IoT \cite{ameer2022attribute}. These authors introduce and compare $\mathrm{HABAC_\alpha}$ (an attribute-based access control model for smart-home IoT) with the EGRBAC (extended generalized role-based access control). The configurations for the role-based approach are mapped with the attribute-based models using user/session, environment, device, operation, and more than one type of attribute. Recently, Sikder et al. introduced a mechanism KRATOS+ for multi-user multi-device access management in Smart home system \cite{sikder2022s}. They implement the idea using four components; user interaction module, backend server, policy manager, and policy execution module. In the user interaction module, the priority management data and device policies are collected. 
This work presents the policy negotiation algorithm and maps the policy to a rule. However, this work is very specific to multi-user shared device environments such as smart homes. 

Relationship-based access control (ReBAC) models \cite{clark2022relog, chakraborty2021feasibility, arora2022higher} have been used to incorporate relations between entities as an access parameter. 
Multilevel relationships are expressed using ABAC models according to this research. Bayreuther and others recently proposed a task planning for a humanoid robot \cite{bayreuther2022bluesky}, which converges to the activity-centric access control \cite{gupta2021towards, mawla2022bluesky} and usage control \cite{park2004uconabc} showing a structure to incorporate policies, objects, modeling framework, architecture and enforcement of the access control system. The authors discuss a decentralized architecture for the policies and task modeling and gain the enforcement of activity-centric and usage-based access control for robot task planning. However, this work lacks the idea of leveraging both models, which is critical for a smart environment. Mawla et al. proposed \cite{mawla2022bluesky} a framework for the activity-centric access control model components to check an activity request. These components fit well to address scenarios that consider activity dependencies and other decision factors. 

Furthermore, several blockchain-based access control solutions are proposed by researchers \cite{tan2021blockchain, han2021blockchain, qin2021lbac, algarni2021blockchain}. Tan et al. propose a blockchain-based access control for the Green Internet of Things (GIoT) for the purpose of saving energy. In this approach, the permission data and identity data are immutable. If we compare this solution to our approach, ACAC is more suitable for scenarios with a large number of devices, a dynamic environment, and supporting dependencies among different activities in smart and collaborative systems. A deep learning-based access control (DLBAC) is proposed by Nobi et al. \cite{nobi2022toward} addresses major limitations of classical access control approaches such as RBAC and ABAC models. This work is significant since it fully automates access control using deep learning. However, it has not been used for large-scale, complex, and dynamic environments due to a lack of accurate access control decisions.
\vspace{-2.5mm}
\section{Conclusion}
\label{conclusion}
In this work, we present a novel activity-centric access control (ACAC) approach for smart and collaborative systems. Considering activity as the prime notion and abstraction to control, we propose an active and object-agnostic security model, which captures the real-time and holistic context of the system to make an activity request decision. Focusing on the dependencies (D) among activities as one of the critical parameters, we formally develop a  family of $\mathrm{ACAC_{D}}$ models supporting activity mutability. We also investigate the chain of dependencies (where dependent activities also can have dependencies) while changing the state of a mutable activity. Resolving chain of dependencies to accommodate the mutability of an activity may be challenging in terms of multiple dependency paths, race conditions and deadlock situations. We explain these challenges and propose potential solutions to deal with those. We also present a prototype implementation of $\mathrm{ACAC_{D}}$ sub-models with a comprehensive smart farming use case reflecting the use of combinations of $\mathrm{ACAC_{D}}$ sub-models and chain of dependencies. Performance is evaluated by the execution time to process many requests with different numbers of  pre-, ongoing, and post-dependent activities' checks and updates.

In the future, we aim to extend this work to a fully mature ACAC model integrating all four authorizations (A), obligations (B), conditions (C), and dependencies (D) parameters. Moreover, our future direction includes developing a formal policy specification language incorporating the chain of dependencies along with other components and analyzing the reachability of incompatible activities as well. Further, a detailed performance evaluation in a real environment having different decision parameters will re-enforce the applicability of the ACAC model in large-scale smart systems.

\bibliographystyle{unsrt}
\bibliography{bibliography}

\end{document}